\documentclass[11pt]{amsart}
\usepackage{graphicx}
\usepackage{float}
\usepackage{amsmath}
\usepackage{mathtools}

\begin{document}
\title[Closed form expression for the Green's function of 
a quantum graph]{Closed form expressions for the Green's 
function of a quantum graph -- a scattering approach}

\author{Tristan Lawrie, Sven Gnutzmann,  Gregor Tanner}

\address{School of Mathematical Sciences, 
University of 
Nottingham,  NG7 2RD, UK}
\email{tristan.lawrie@nottingham.ac.uk}
\vspace{10pt}

\begin{abstract}
    In this work we present a three step procedure for 
    generating a closed form expression of the Green's 
    function on both closed and open finite quantum graphs 
    with general self-adjoint matching conditions. 
    We first generalize and simplify the approach by Barra 
    and Gaspard [Barra F and Gaspard P 2001, 
    Phys.~Rev.~E {\bf 65}, 016205] and then discuss
    the validity of the explicit expressions.
    For compact graphs, we show that the explicit expression is
    equivalent to the spectral decomposition as a sum over poles
    at the discrete energy eigenvalues with residues that contain projector
    kernel onto the corresponding eigenstate.\\
    The derivation of the Green's function is based on the scattering approach, in 
    which 
    stationary solutions are constructed by treating each 
    vertex or subgraph as a scattering 
    site described by a scattering matrix.
    The latter can then be given in a simple
    closed form from which 
    the Green's function is derived.\\ 
    The relevant scattering matrices 
    contain inverse operators which are not well defined 
    for 
    wave numbers at which bound states in the continuum 
    exists. It is shown that the singularities in the 
    scattering matrix related to these bound states or 
    perfect scars  can be regularised. 
    Green's functions or scattering matrices can then 
    be expressed as a sum of a 
    regular and a singular part where the singular part 
    contains the projection kernel onto the perfect scar.

\textit{Keywords}: Quantum Graphs, Green's functions, Wave 
scattering.
\end{abstract}

\maketitle


\section{Introduction}

Quantum graphs as
metric graphs endowed with a 
Schr\"odinger operator and related similar models
have a long history in mathematics, physics and theoretical
chemistry
\cite{Pauling36, RS53, Coulson54, Montroll70, Roth83, 
Alexander83, Below88}. 
Due to the simplicity of the model and the 
richness of properties and effects it can represent, quantum 
graphs have grown into an important tool in physics 
and mathematics.
In spectral theory, they allow for a rigorous treatment of 
topics that are usually related to the study of (self-
adjoint) partial differential operators, see \cite{BK13} for 
an introduction and overview. 
The scattering approach to quantum graphs was introduced
in 1997 by Kottos and Smilansky \cite{KS97}
and led to a wide range of applications in quantum chaos,
see \cite{GS06} for an overview.
In this approach, the graph vertices are treated as 
scattering sites from which stationary solutions (energy
eigenstates) are constructed. 
This approach has also been 
used for many physical applications beyond quantum chaos, 
including meta-material design \cite{LTC22}, modelling the 
vibrations of coupled  plates \cite{BCT18}, as well as in 
formulating quantum random walks \cite{Kempe03, Tan06} and 
quantum search algorithms \cite{HT09}. One advantage of the 
scattering approach is that eigenvalue conditions can be 
written in terms of a secular equation involving the 
determinant of a unitary matrix of finite dimension $N$, 
where $N$ typically equals twice the number of edges on the 
graph. Similarly, the scattering matrix of an open quantum 
graph can be given in terms of a closed form expression 
involving finite dimensional matrices of size $N$ \cite{KS00, 
BG01}.

In 2001, Barra and Gaspard \cite{BG01}
used the scattering approach to express the Green's
function of a quantum graph as a sum over trajectories
in the spirit of semiclassical quantum mechanics. 
At the time, it was not yet clear within the physics community
what scattering matrices are connected to matching conditions 
related to a well-defined
self-adjoint Schr\"odinger operator on the metric graph.
We generalize and simplify the approach \cite{BG01}
by using a simple three step procedure 
that leads to the Green's function for general
self-adjoint matching conditions for closed and open graphs
with a finite number of edges.
This directly provides a number of closed form 
expressions that,
to the best of our knowledge, have not been given before 
(though implied in \cite{BG01}, see also \cite{SCL03},
where closed form expressions are given for a few simple examples). 
These closed forms are of 
great practical advantage when dealing with explicit graphs
as they sum all relevant trajectories. Moreover, they are
the starting point of an analysis of the validity and
convergence of Green's function when expressed as a sum over 
trajectories.
We thus hope 
to provide a more straightforward way of 
computing Green's functions on graphs. This could lead to 
helpful insight into the growing literature on applications 
for Green's functions on graphs that often require
relatively cumbersome sums over trajectories,
see \cite{ASVCL16, AS18, SAB21} and references 
therein. 

We also discuss in some detail cases where the sum over 
trajectories fails to converge while closed form expressions
may be regularized.
Indeed, when evaluating the scattering matrix 
on open graphs, such as 
those used in the construction of the Green's function, one 
must take great care at frequencies corresponding to bound 
states in the continuum. 
These states vanish necessarily on the scattering leads 
and potentially lead to singular behaviour when 
considering the Green's function. 
Scarring of eigenfunctions is a well-known semiclassical 
phenomenon in more general systems \cite{Heller84}. 
It has been known since the work of Schanz and 
Kottos \cite{SK03} that quantum graphs 
allow for a much stronger scarring mechanism than in more 
general wave systems. These so-called perfect scars 
are non-vanishing only on a finite subset of the edges and vanish exactly
in the remainder of the graph. They
are easily constructed, 
for example, in certain quantum graphs with standard (Neumann-
Kirchhoff) vertex 
matching conditions. For open graphs, bound states in the continuum are 
an example of perfect scars.
Perfect scars
lead to singularities in some inverted matrices
that are used in the construction of scattering matrices
and Green's function and this implies non-convergence of 
the related
sums over trajectories at the corresponding wave number.
We will 
explain that both the scattering matrix and the Green's 
function (outside the domain of the perfect scar) 
stay regular at these frequencies and give suitable 
regularised equations.
These regularized expressions may be of practical
importance even if there are no perfect scars on the graph.
This is due to the far more generic phenomenon of almost 
perfect scars which was also described in 
\cite{SK03}. These are states where the conditions 
for a perfect scar on a subgraph are fulfilled up to 
small terms
leading to states which are only slightly coupled to outgoing channels. 
In the scattering matrix, almost perfect scars lead to
what is known as 
topological resonances \cite{GSS13,verdiere18}.
In this context, a simplified variant of the regularization 
scheme we describe here has been used to
derive the tails in the distribution of resonance widths
\cite{GSS13}.

The paper is structured as follows: In Sec.~\ref{sec:qgraphs}, 
the scattering representation is introduced 
for both closed and open quantum graphs. 
In Sec.~\ref{sec:Green}, a three step procedure for 
generating a closed form expression for the Green's function 
is introduced via the scattering approach. The expression is 
given generally for both closed and open quantum graph. It is 
assumed that the graph scattering matrix is non-singular and 
well defined.
In Sec.~\ref{sec: bound states}, a formal definition is given 
for a scar state in terms of the quantum map. It is shown 
that the block component of the quantum map that refers to 
the compact portion of the graph is non-invertible. 
It is shown through a regularization of the scattering 
approach, that the full solution is indeed regular as it is 
evaluated within a reduced space. Further analysis of the 
scattering states for eigenenergies approaching a scar state 
are investigated \ref{AppB}.
In Sec.~\ref{sec:examples}, we generate the scattered states 
and the corresponding Green's function in the presence of 
scars for two examples, the open lasso, and the open star 
graph. We finally conclude this work in Sec.~\ref{sec:conclusion} with a 
brief summary and outlook.

\section{The scattering approach for quantum graphs}\label{sec:qgraphs}
To construct a quantum graph, we first consider a metric graph
\(\mathcal{G}(\mathcal{V},\mathcal{E},L)\).
Here, \(\mathcal{E}\) is the set of edges,
\( \mathcal{V} \) the set of vertices,
and \( L = \{\ell_e :  e \in \mathcal{E} \}  \) is the graph metric 
containing 
a set of edge lengths which are either real positive \(\ell_e>0\) or infinite.
The set of edges with finite length will be called the set of 
bonds \(\mathcal{B}\) and the set of edges 
with infinite length will be called the set of leads \(\mathcal{L}\).
We consider two types of finite graphs:\\ 
i. Closed compact graphs where all edges are bonds and the number of edges
\(N_{\mathcal{E}}= |\mathcal{E}|\) is finite. Here, both ends of each edge 
are connected 
to a vertex.\\
ii. Open scattering graphs which consist of a compact graph with the 
addition 
of a finite set of leads. The leads are connected to a single vertex at one 
end.
One may write the edge set as a union 
\(\mathcal{E}=\mathcal{L} \cup \mathcal{B} \).
With \(N_{\mathcal{L}}=|\mathcal{L}|\) and 
\(N_{\mathcal{B}}=|\mathcal{B}|\), one has
\(N_{\mathcal{E}}=N_{\mathcal{B}}+N_{\mathcal{L}}\). \\
For each bond \(e\in \mathcal{B} \), we use a coordinate 
\(x_e \in [0, \ell_e]\) 
with some (arbitrary but fixed) choice of direction. 
The coordinate defines a position on an
edge such that \(x_e=0\) and \(x_e= \ell_e\) correspond to the vertices 
connected 
by the bond. For a lead \(e\in \mathcal{L}\), coordinates 
\(x_e \in [0,\infty) \)
are defined such that \(x_e=0\) corresponds to the vertex where the lead is 
attached.
For each edge \(e\), we refer to the directed edges as \(e_s\)
with \(s=\pm\) indicating the direction in which 
\(x_e\)
increases ($s=+$) or decreases ($s=-$). 
A point on the graph is a pair \(\mathbf{x}=(e, x_e)\) of an edge
and a coordinate.

The metric graph is turned into a quantum graph by adding a 
Schr\"odinger operator \(\hat{H}\) which requires a set of boundary 
conditions on the graph vertices in order to become a self-adjoint 
problem. For this, we consider the Hilbert space
\( L^2(\mathcal{G})\equiv \bigoplus_{e \in E} L^2([0,\ell_e]) \) of square 
integrable complex-valued functions 
\(\boldsymbol{\Phi}(\mathbf{x})=\{\phi_e(x_e) \}_{e \in \mathcal{E}}\) 
and define 
\begin{equation}
\label{operator}
    \left[\hat{H}\boldsymbol{\Phi}(\mathbf{x})\right]_{e} 
    = - \frac{d^2}
    {dx_e^2}\phi_{e}(x_{e})+ V_e(x_e)\phi_{e}(x_{e})
\end{equation}
with a potential \(\boldsymbol{V}(\mathbf{x})=
\{V_e(x_e)\}_{e \in \mathcal{E}} \), that is, a real valued scalar 
function defined on \(\mathcal{G}\).  
We will only consider free Schr\"odinger operators,
that is, negative Laplacians, where \(\boldsymbol{V}(\mathbf{x})=0\).
To ensure that the second derivative is well defined and
square integrable, one needs to restrict the domain 
of \(\hat{H} \) 
to an appropriate Sobolev space. Apart from this 
standard restriction, the domain of \( \hat{H}\)
has to be further specified by appropriate boundary conditions 
at each vertex $v$ in order for \(\hat{H}\) to define a 
self-adjoint operator.
%
According to a theorem by
Kostrykin and Schrader \cite{KSchr99}, the most general such boundary 
conditions 
at the vertex \(v\) may be written in the form
\begin{equation} \label{bc}
    \sum_{\tilde{e}} 
    \boldsymbol{A}
    _{e\tilde{e}} \phi_{\tilde{e}}(0) 
    +
    \boldsymbol{B}
    _{e\tilde{e}} 
    \frac{d\phi_{\tilde{e}}}{dx_{\tilde{e}}}(0) =0
\end{equation}
for any \(e\) connected to $v$ and the sum extends over 
edges \(\tilde{e}\) connected to \(v\).
(We assumed here for simplicity that $x_e =0$ at the 
vertex for each edge $e$ 
connected to $v$.) The complex 
coefficients \(
\boldsymbol{A}
_{e\tilde{e}}\) and 
\(
\boldsymbol{B}
_{e\tilde{e}}\) refer to the 
elements $e\tilde{e}$ of 
two square matrices \(\boldsymbol{A}\) and \(\boldsymbol{B}\) of 
dimension \(d_v\), 
the number of edges connected to $v$. In \cite{KSchr99}, it was proven 
that the 
matching conditions preserve self-adjointness if and only if two 
conditions are 
satisfied. 
First, the set 
of equations need to be independent which means that the rectangular 
\(d_v\times 2 d_v\) matrix \( 
\left(\boldsymbol{A},\boldsymbol{B}\right)\), i.e.\ 
$\boldsymbol{A}$ and $\boldsymbol{B}$ being horizontally 
stacked, must have full 
rank \(d_v\). Second, the product 
\(\boldsymbol{AB}^\dagger= 
\boldsymbol{BA}^\dagger\) is a Hermitian matrix. 
The matrices \(\boldsymbol{A}\) and \(\boldsymbol{B}\) may 
be chosen independently 
for each vertex and we will often write 
\(\boldsymbol{A}^{(v)}\) and 
\(\boldsymbol{B}^{(v)}\) to indicate the vertex where 
these matrices act.

The self-adjointness of \(\hat{H}\) implies a unitary 
evolution
of the time-dependent Schr\"odinger equation 
\( i \frac{d}{dt} \boldsymbol{\Phi}(t) =\hat{H} 
\boldsymbol{\Phi}(t)\). 
The stationary solutions 
\(\boldsymbol{\Phi}(t)= e^{-i E t} \boldsymbol{\Psi} \)  
satisfy the (homogeneous) eigenproblem
\begin{equation}
\label{homogenious equation}
    \left[\left(E-\hat{H}\right)\boldsymbol{\Psi}
    (\mathbf{x})\right]_e =
    \left(E + \frac{d^2}{dx_e^2} \right)\psi_e(x_e)
    =0\ .
\end{equation}
Here, \(E\) 
is the energy. It implies furthermore that solutions
to \eqref{homogenious equation} only exist for real values of \(E\) and
the set of all such (generalized) eigenvalues forms the spectrum of 
\(\hat{H}\).
In the remainder we will only consider the positive part of their spectrum
and write
\(E=k^2>0\) with the wave number \(k>0\).
In the following constructions, the energy appears
as a variable that is not restricted to the spectrum.

Any solution to equation (\ref{homogenious equation}) fulfilling the 
prescribed 
boundary conditions at the vertices is expressed as a superposition of 
counter 
propagating plane waves, that is,
\begin{equation}
\begin{split}
\psi_e(x_e)=& a^{\mathrm{in}}_{e_-}e^{-ikx_e}+a^{\mathrm{out}}_{e_+}
    e^{ikx_e}\\
    =& a^{\mathrm{out}}_{e_-}e^{-ik(x_e-\ell_e)}+
    a^{\mathrm{in}}_{e_+}e^{ik(x_e-\ell_e)}\\
    =&
    a^{\mathrm{in}}_{e_-}e^{-ikx_e}+
    a^{\mathrm{in}}_{e_+}e^{ik(x_e-\ell_e)}
    \ .
\end{split}
\label{plane_wave_solution}
\end{equation}
Here, \(a^{\mathrm{in/out}}_{e_\pm}\) is the complex wave amplitude on 
edge $e$ 
propagating in the direction of increasing (\(+\)) or decreasing (\(-\)) 
\(x_e\), 
heading \(\mathrm{in}\) or \(\mathrm{out}\) of a vertex. If \(e\) is a 
lead only 
the amplitudes \(a^{\mathrm{in/out}}_{e_\pm}\) at \(x_e=0\) are used.\\
Introducing the \(2N_B\)-dimensional diagonal length matrix 
\begin{equation}
    \mathbf{L}_{\tilde{e}_{\tilde{s}} e_s}=
    \delta_{e\tilde{e}}\delta_{s\tilde{s}} \ell_e
\end{equation}
(where each edge length appears twice)
the bond wave amplitudes can be mapped to one another 
by the diagonal 
square \(2N_B\)-dimensional
matrix 
\begin{equation}
    \mathbf{T}(k)=e^{i k\mathbf{L}}
\end{equation}
that takes account of the phase difference 
between wave 
amplitudes across all bonds, that is,
\begin{equation}
    \mathbf{a}^{\mathrm{in}}_{\mathcal{B}}= \mathbf{T}(k)\  
    \mathbf{a}^{\mathrm{out}}_{\mathcal{B}}\ .
    \label{bond_transport}
\end{equation}
Here, \(\mathbf{a}^{\mathrm{in/out}}_{\mathcal{B}}\) refers to the 
\(2N_{\mathcal{B}}\) vector
of plane wave coefficients on the directed bonds. \\
In addition, the graph wave amplitudes can be mapped onto one another 
across the 
vertices by taking account of the imposed vertex boundary conditions. For 
this one 
writes the matching conditions
at a given vertex \(v\) in the form of a $d_v \times d_v$ vertex scattering 
matrix 
$\boldsymbol{\Sigma}^{(v)}$, that is,
\begin{equation}
\mathbf{a}^{(v),\mathrm{out}}= \boldsymbol{\Sigma}^{(v)} 
\mathbf{a}^{(v),\mathrm{in}}
\end{equation}
where \(\mathbf{a}^{(v),\mathrm{in/out}}\) are \(d_v\) dimensional vectors 
that 
collect all incoming/outgoing amplitudes of plane waves on the edges \(e\) 
in the 
neighborhood of vertex \(v\). With the prescribed boundary conditions given 
in 
Eq.\ (\ref{bc}), $\boldsymbol{\Sigma}^{(v)}$ takes on the form
\begin{equation}
    \boldsymbol{\Sigma}^{(v)}(k) = -\left(\boldsymbol{A}^{(v)}+ i k 
    \boldsymbol{B}^{(v)}\right)^{-1} 
    \left(\boldsymbol{A}^{(v)} - ik \boldsymbol{B}^{(v)}\right).
    \label{general_vertex_scattering}
\end{equation} 
For real \(k\) (\(E>0\)), this is a well-defined unitary matrix
due to the conditions on \(\boldsymbol{A}^{(v)}\) and 
 \(\boldsymbol{B}^{(v)}\)
which imply that \(\boldsymbol{A}^{(v)}+ ik\boldsymbol{B}^{(v)} \)
is invertible.
Note,  
however, that neither
\(\boldsymbol{A}^{(v)}\) 
nor 
\(\boldsymbol{B}^{(v)}\) 
need to be invertible by themselves
(in general neither is) and
one needs to take care at 
\(k = 0\), for instance, where it remains well defined as a limit.
Another consequence is that the explicit dependence on \(k\) may drop
for some choices of matching conditions. Indeed, this
is the case for the 
so-called Neumann-Kirchhoff matching conditions
most widely used in the literature \cite{BK13,KS97,GS06}.
They require continuity of the wave function at the vertex
\(\phi_e(0)=\phi_{\tilde{e}}(0)\) (for any \(e\) and \(\tilde{e}\) 
connected to \(v\)) and 
a vanishing sum of outward derivatives on the edges connected to 
this vertex \(\sum_{e} \frac{d\phi_e}{d x_e}(0)=0\) (where the sum is over 
all edges connected to \(v\)).
This yields
\begin{equation}\label{Kirchoff-Neumann BC}
    \boldsymbol{\Sigma}^{(v), \mathrm{NK}}=  - \boldsymbol{\mathbb{I}} +
    \frac{2}{d_v}\boldsymbol{\mathbb{E}}_{d_v},
\end{equation}
where \(\mathbb{I}\) is the identity matrix and \(\mathbb{E}_{d_v}\) is the 
matrix 
of dimension \(d_v\) with all entries equal to one. 

It is worth noting that in the physics literature including
\cite{BG01}, the stationary problem is often defined on a quantum graph 
by prescribing arbitrary unitary matrices \(\Sigma^{(v)}\) at 
the vertices \(v\). 
While this does in general not define an operator in a
Hilbert space (self-adjoint or not) this is of obvious value for 
an effective description of a
physical system if appropriate caution is used. For instance, one should not 
expect eigenstates to be orthogonal and time-dependent solutions obtained by
superposition may not preserve probability (the norm).
In some applications that focus on spectral properties, for instance many 
applications in quantum chaos, these issues are not physically relevant,
see \cite{GS06} and many references therein. Moreover, they may be given
physical meaning by assuming that a vertex stands for a hidden part of the 
system, 
such as a scattering region, thus also `hiding' parts of the Hilbert space.
In the following, we will assume that scattering matrices are of the form
\eqref{general_vertex_scattering} that ensures a self-adjoint operator.  
Most of our results remain valid if arbitrary scattering matrices
are prescribed as long as they do not depend explicitly on the wave number.

One may combine all vertex scattering matrices into a single 
(directed) edge scattering matrix \(\boldsymbol{\Sigma}\), such that
\begin{equation}
  \mathbf{a}^{\mathrm{out}}=  \boldsymbol{\Sigma} \; 
  \mathbf{a}^{\mathrm{in}}.
  \label{graph_scattering}
\end{equation}
Here, \(\mathbf{a}^{\mathrm{in/out}}\) is a \(2N_B+N_L\) dimensional vector 
that 
collects all the incoming/outgoing 
amplitudes for all graph bonds and leads.
The scattering matrix elements are expressed in terms of the individual 
vertex 
scattering matrices $\boldsymbol{\Sigma}^{(v)} $, such that, after ordering 
the directed edges in an appropriate way, 
\begin{equation}\label{hatSigma}
    \mathbf{\Sigma}= \mathbf{\Pi} 
    \begin{pmatrix}
       \boldsymbol{\Sigma}^{(1)} & 0& \dots& 0\\
       0 & \boldsymbol{\Sigma}^{(2)} & \dots & 0 \\
       \vdots & \vdots & \ddots & \vdots \\
       0 & 0 & \dots & \boldsymbol{\Sigma}^{(N_V)} 
    \end{pmatrix} = \mathbf{\Pi}\, \mathbf{\hat{\Sigma}}\, .
\end{equation}
Here, \(\mathbf{\Pi}\) is a permutation matrix that interchanges the two 
directions on a given edge with matrix elements given as
\begin{equation}
    \boldsymbol{\Pi}
    _{\tilde{e}_{\tilde{s}} e_s}=
    \delta_{\tilde{e} e} \delta_{\tilde{s} (-s)}\ .
    \label{directed_edge_permutation}
\end{equation}

\subsection{Compact quantum graph eigenstates in the 
scattering representation}

In the case of a compact quantum graph, we have 
\(\mathbf{a}^{\mathrm{in/out}}_{\mathcal{B}} \equiv 
\mathbf{a}^{\mathrm{in/out}} 
\). 
The two relations \eqref{bond_transport} and \eqref{graph_scattering}  
combine 
to give one condition, 
\begin{equation}
    \mathbf{a}^{\mathrm{in}} = \mathbf{U}(k)\  \mathbf{a}^{\mathrm{in}}, 
    \label{quantization}
\end{equation}
forming the \(2N_{\mathcal{B}}\) dimensional quantum map
\begin{equation}
    \mathbf{U}(k)= \mathbf{T}(k) \mathbf{\Sigma}(k),
    \label{quantum_map}
\end{equation}
where we stress that the edge scattering matrix 
\(\mathbf{\Sigma}(k)\) can be $k$ 
dependent. Non-trivial solutions 
to \eqref{quantization} exist for wave numbers \(k\) for which the quantum 
map 
\(\mathbf{U}\) has a 
unit eigenvalue, that is, for wave numbers that satisfy the secular equation
\begin{equation} \label{secular}
\xi(k)\equiv \det\left(\mathbb{I}-{\bf U}(k) \right)=0\ .
\end{equation}
The positive (discrete) energy spectrum of the quantum graph 
corresponds one-to-one to the zeros of $\xi(k)$ with \(k>0\) 
\cite{KS97,KSchr99,BE09}. 
The corresponding eigenstates can be obtained from \eqref{quantization}.

\subsection{Scattering states on open quantum graphs}

Let us consider the positive energy states for open 
quantum graphs next.
Generically, these consist of an \(N_{\mathcal{L}}\)-fold degenerate 
continuum of scattering states. Physically, the 
\(N_{\mathcal{L}}\)-fold degeneracy is obvious from
the ability to choose \(N_{\mathcal{L}}\) independent incoming plane 
waves along the leads. To describe the scattering states, let us 
write the unitary edge scattering matrix in block form
\begin{equation}
    \mathbf{\Sigma}(k)= 
    \begin{pmatrix}
       \mathbf{\Sigma}(k)_{\mathcal{LL}} & \mathbf{\Sigma}(k)_{\mathcal{LB}}\\
       \mathbf{\Sigma}(k)_{\mathcal{BL}} & \mathbf{\Sigma}(k)_{\mathcal{BB}}
    \end{pmatrix}= 
    \begin{pmatrix}
    \mathbb{I} & 0\\
    0 & \mathbf{\Pi}
    \end{pmatrix} 
     \mathbf{\hat{\Sigma}}(k) ,
\end{equation}
where the block-indices \(\mathcal{B}\) and \(\mathcal{L}\) refer to 
\(2 N_{\mathcal{B}}\) 
directed bonds and \(N_{\mathcal{L}}\) leads.
In the second equality, we have expressed this explicitly
in terms of the matrix \(\mathbf{\hat{\Sigma}}(k)\) defined in 
\eqref{hatSigma}
which is block-diagonal in the vertex scattering matrices and the 
permutation 
matrix \(\mathbf{\Pi}\) that interchanges the two directions for any two 
bonds
as defined in \eqref{directed_edge_permutation}.
For an open quantum graph, \(\mathbf{\Pi}\) only acts on bonds. 
Analogously to the compact case in Eq.\ (\ref{quantum_map}), we introduce 
the unitary quantum map for
an open graph, again expressed in block form, 
\begin{equation}
    \mathbf{U}(k)\equiv
    \begin{pmatrix}
       \mathbf{U}(k)_{\mathcal{LL}} & \mathbf{U}(k)_{\mathcal{LB}}\\
       \mathbf{U}(k)_{\mathcal{BL}} & \mathbf{U}(k)_{\mathcal{BB}}
    \end{pmatrix}
    =
    \begin{pmatrix}
       \mathbf{\Sigma}(k)_{\mathcal{LL}} & 
       \mathbf{\Sigma}(k)_{\mathcal{LB}}\\
       \mathbf{T}(k)\mathbf{\Sigma}(k)_{\mathcal{BL}} & \mathbf{T}
       (k)\mathbf{\Sigma}(k)_{\mathcal{BB}}
    \end{pmatrix}\ .
    \label{quantum_map_scattering}
\end{equation}

The scattering states are spanned by the \(N_L\)-dimensional
vector \(\mathbf{a}^{\mathrm{in}}_{\mathcal{L}}\) of incoming plane wave 
amplitudes on the leads.
The outgoing amplitudes \(\mathbf{a}^{\mathrm{out}}_{\mathcal{L}}\)
and the incoming amplitudes on the directed bonds
\(\mathbf{a}^{\mathrm{in}}_{\mathcal{B}}\) then result from solving
the set of linear equations
\begin{equation}
    \begin{pmatrix}
       \mathbf{a}(k)^{\mathrm{out}}_{\mathcal{L}}\\
       \mathbf{a}(k)^{\mathrm{in}}_{\mathcal{B}}
    \end{pmatrix}
    =
    \begin{pmatrix}
       \mathbf{U}(k)_{\mathcal{LL}} & \mathbf{U}(k)_{\mathcal{LB}}\\
       \mathbf{U}(k)_{\mathcal{BL}} & \mathbf{U}(k)_{\mathcal{BB}}
    \end{pmatrix}
     \begin{pmatrix}
       \mathbf{a}^{\mathrm{in}}_{\mathcal{L}}\\
       \mathbf{a}(k)^{\mathrm{in}}_{\mathcal{B}}
    \end{pmatrix}
    \label{scattering_linear_equations}
\end{equation}
which follows again from \eqref{bond_transport} and \eqref{graph_scattering}.
Solving these equations, one obtains for 
the outgoing amplitudes on the leads
\begin{equation}
    \mathbf{a}(k)^{\mathrm{out}}_{\mathcal{L}}= \boldsymbol{\sigma}(k) 
    \mathbf{a}^{\mathrm{in}}_{\mathcal{L}}
    \label{scattering_amplitudes_leads}
\end{equation}
where the unitary   
graph scattering matrix is given as
\begin{equation}
    \boldsymbol{\sigma}(k) = \mathbf{U}(k)_{\mathcal{LL}}+ 
    \mathbf{U}(k)_{\mathcal{LB}}
    \frac{\mathbb{I}}{\mathbb{I}-\mathbf{U}(k)_{\mathcal{BB}}}
    \mathbf{U}(k)_{\mathcal{BL}}\ .
    \label{scattering_matrix}
\end{equation}
The plane wave amplitudes on the directed bonds 
can be expressed as 
\begin{equation}
    \mathbf{a}(k)^{\mathrm{in}}_{\mathcal{B}}
    = \boldsymbol{\rho}(k) \mathbf{a}^{\mathrm{in}}_{\mathcal{L}}
    \label{rho_definition}
\end{equation}
with the rectangular \(2N_{\mathcal{B}} \times N_{\mathcal{L}}\) matrix
\begin{equation}
    \boldsymbol{\rho}(k)= \frac{\mathbb{I}}{\mathbb{I}-
    \mathbf{U}(k)_{\mathcal{BB}}}
    \mathbf{U}(k)_{\mathcal{BL}}\ .
    \label{scattering_state}
\end{equation}
The scattering matrix \(\boldsymbol{\sigma}(k)\) is related to the
matrix \(\boldsymbol{\rho}(k)\) via
\begin{equation}
    \boldsymbol{\sigma}(k) = \mathbf{U}(k)_{\mathcal{LL}} + 
    \mathbf{U}(k)_{\mathcal{LB}}\ \boldsymbol{\rho}(k)\ .
    \label{sigma_in_terms_of_rho}
\end{equation}
We now have the required mathematical language for constructing Green's 
functions on quantum graphs.

One may rightfully question whether the matrix 
\( \mathbb{I}-\mathbf{U}(k)_{\mathcal{BB}}\) can always be inverted as 
required in equations \eqref{scattering_matrix} and 
\eqref{scattering_state}. 
This is related to the existence of bound 
states in the continuum (a pure point spectrum in mathematical terms).
In the absence of such bound states \(\mathbf{U}(k)_{\mathcal{BB}}\) 
does not have a unit eigenvalue and the expression is valid for all wave 
numbers \(k>0\). We will return to the discussion of this expression in the 
presence of bound states, also known as perfect scars, 
later in Sec.~\ref{sec: bound states}.

\section{The scattering approach to the Green's function}
\label{sec:Green}

The Green's function may be considered as the integral kernel
of the resolvent operator
\( (E - \hat{H})^{-1}\) which has singularities at the spectrum
of \(\hat{H}\). It has poles at the discrete spectrum and a branch cut along the 
continuous spectrum. 

For a given (complex) energy \(E = k^2\) and two points \(\mathbf{x}=(e,x_e)\)
and \(\mathbf{x'}=(e',x'_{e'})\) on a quantum graph, the Green's function 
\(G(\mathbf{x},\mathbf{x'},E)\)  satisfies the inhomogeneous equation
\begin{equation}
    \left(E-\hat{H}\right) G(\mathbf{x},\mathbf{x'},E)
    = \delta(\mathbf{x},\mathbf{x}')
    \equiv
    \begin{cases}
    \delta(x_e-x'_{e'}) & \text{if \(e=e'\)}\\
    0 & \text{if \(e \neq e'\)}
    \end{cases}, 
    \label{Definition_Green}
\end{equation}
where \(\hat{H}\) acts on \(\mathbf{x}\).
The solution of this differential equation \eqref{Definition_Green} with given self-adjoint matching 
conditions at the vertices
 may not be unique or not exist at all. 
The latter happens when the energy \(E\) 
belongs to the discrete real eigenvalue spectrum.
For complex energies with a non-vanishing imaginary part, one can always find a 
unique
square integrable solution and this then coincides with the integral kernel
of the resolvent operator.
The relation to the resolvent operator gives rise to the symmetry
\begin{equation}
    G(\mathbf{x},\mathbf{x'};E)= G(\mathbf{x'},\mathbf{x};E^*)^*\ .
    \label{GF_symmetry}
\end{equation}
We focus on the Green's function 
\(G_+(\mathbf{x},\mathbf{x'},E)\equiv
G(\mathbf{x},\mathbf{x'},E_+) \) with positive real and imaginary parts: 
\(E_{+} =k_+^2= E_{r}+ iE_{i}\) with \(0<E_{r} \in 
\mathbb{R}\)  and \(0 < E_{i} \in \mathbb{R}\).
For real energies that are not in the (discrete or continuous) eigenvalue spectrum,
we allow the imaginary part to vanish, that is, \(E_i=0\), as the Green's function is well 
defined in that case.
Solutions at real energies in the continuous spectrum require the limit \(E_i \to 
0^+ \)  which is always implied.
If \(E_r\) belongs to the discrete 
eigenvalue spectrum, the Green's function has a pole 
\(G(\mathbf{x},\mathbf{x'};E) \sim \frac{P(\mathbf{x},\mathbf{x'})}{E_i}\) 
(with a non-vanishing function 
\(P(\mathbf{x},\mathbf{x'})\)) preventing 
the limit \(E_i \to 
0^+ \) to exist.
For brevity we write $E = E_{+}$ and $k = k_{+}$ during the following 
derivations.

To construct the Green's function, we exploit the fact that for all 
\(\mathbf{x} \neq \mathbf{x'}\) the solutions to equation (\ref{Definition_Green}) 
are solutions to the homogeneous wave equation in (\ref{homogenious equation}). 
This allows one to express the solutions again as a linear superposition of 
counter propagating plane waves as express in (\ref{plane_wave_solution}). The set 
of unknown coefficients are then chosen to satisfy the imposed vertex boundary 
conditions as well as the appropriate boundary conditions at the delta function 
excitation \(\mathbf{x} = \mathbf{x'}\). This procedure is detailed via a scattering approach 
in the following. 

\subsection{Construction of the Green's function for compact graphs} 
\label{sec:green_compact}
\begin{figure}[H]
\centering
\includegraphics[width = 0.85\textwidth]{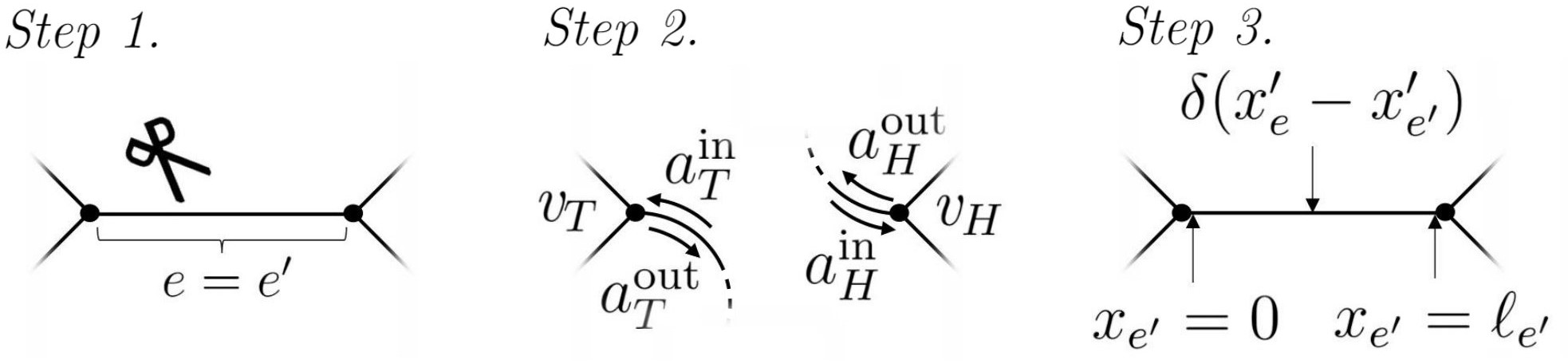}
\caption{This three step procedure is described in detail below.
\label{fig:Green's function Graph}}
\end{figure}

The Green's function on a graph can be constructed in a three step procedure as 
illustrated in Fig.~\ref{fig:Green's function Graph}.

\begin{itemize}

    \item[\emph{Step 1.}] Define the graph and the coordinate of the delta 
    function excitation \(\mathbf{x'}=(e',x'_{e'})\). The delta function acts as a 
    source which we model by creating an auxiliary open scattering graph by 
    ``cutting out'' the excited edge \(e'\) and replacing it with two auxiliary 
    leads.
    
    \item[\emph{Step 2.}] Treat the auxiliary graph as a scattering site and 
    construct a lead scattering matrix for energy $E_+$. This allows one to 
    determine the two outgoing lead wave amplitudes in terms of the two incoming 
    wave amplitudes which are free parameters.
    
    \item[\emph{Step 3.}] Take the scattering solution on the auxiliary leads at 
    distances \(x'_{e'}\)
    and \( \ell_{e'}-x'_{e'}\) from the vertices and "glue" these solutions 
    together such that the differential equation \eqref{Definition_Green} is 
    satisfied yielding a Dirac \(\delta\)-function at the position 
    \(\mathbf{x}'\).
    This determines all free parameters and results in the Green's function 
    \(G(\mathbf{x},\mathbf{x'};E_+)\).
\end{itemize}
Let us now go through these steps in detail:\\
\emph{Step 1.}
Consider a compact quantum 
graph \(\mathcal{G}(\mathcal{V},\mathcal{E},L)\) 
as defined in Sec.\ \ref{sec:qgraphs} which we wish to excite with a delta 
function at location \(\mathbf{x'}=(e',x'_{e'})\in \mathcal{G}\). Let us denote 
the vertex at $x_{e'} = 0$ as the 'tail' vertex $v_{T}$ and the vertex at $x_{e'} 
= l_{e'}$ as the 'head' vertex $v_{H}$. We begin by cutting the excited edge $e'$ and 
replacing it by two leads attached at $v_{T}$ and $v_{H}$, respectively, thus creating the auxiliary open scattering graph 
\(\mathcal{G}_{\mathrm{aux}, e'}=\mathcal{G}_{\mathrm{aux}, e'}(\mathcal{V},
\mathcal{E}_{\mathrm{aux, e'}}, L_{\mathrm{aux, e'}} )\), where 
\(\mathcal{E}_{\mathrm{aux, e'}} =
\mathcal{L}_{\mathrm{aux},e'} \cup (\mathcal{B} \setminus \{e'\})\) and
\( L_{\mathrm{aux},e'}= L \setminus\{ \ell_{e'}\} \). The coordinates on the leads 
are set to be $x_{T} = x_H = 0$ at the vertices $v_{T}$ and $v_{H}$, respectively. 
On each lead, the solutions are defined as
\begin{equation}
    \begin{split}
        & \psi_{T}(x_T) = a_{T}^{\mathrm{in}} e^{-ik_+x_{T}} + 
        a_{T}^{\mathrm{out}} 
        e^{ik_+x_{T}}, \\
        & \psi_{H}(x_H) = a_{H}^{\mathrm{in}} e^{-ik_+x_{H}} + 
        a_{H}^{\mathrm{out}} 
        e^{ik_+x_{H}}.
    \end{split}
\end{equation}
\emph{Step 2.} Next, we construct the scattering states on the auxiliary graph. 
The quantum map of the auxiliary graph
can then be written in the form Eq.\
\eqref{quantum_map_scattering} and only differs from the quantum map
\eqref{quantum_map} of \(\mathcal{G}\) by excluding the rows corresponding to the 
excited edge $e'$. 
The wave amplitudes on the two leads are mapped from incoming to outgoing wave 
amplitudes by the graph scattering matrix $\boldsymbol{\sigma}(k_+)$ as defined in
\eqref{scattering_amplitudes_leads} with matrix elements
\begin{equation} \label{qscatter}
    \begin{pmatrix}
        a^{\mathrm{out}}_{\mathrm{H}}\\
        a^{\mathrm{out}}_{\mathrm{T}}
    \end{pmatrix}
    =
    \begin{pmatrix}
        \sigma(k_+)_{HH} & \sigma(k_+)_{HT} \\
        \sigma(k_+)_{TH} & \sigma(k_+)_{TT}
    \end{pmatrix}
    \begin{pmatrix}
        a^{\mathrm{in}}_{\mathrm{H}}\\
        a^{\mathrm{in}}_{\mathrm{T}}
    \end{pmatrix}.
\end{equation}
The incoming wave amplitudes 
\(a^{\mathrm{in}}_{\mathrm{H}}\)
and \(a^{\mathrm{in}}_{\mathrm{T}}\) are at this stage 
free parameters. \\
\emph{Step 3.}
We project the set of scattering solutions from the 
auxiliary graph onto the 
original graph by cutting the leads \(\mathrm{H}\) and 
\(\mathrm{T}\) at 
\(x_{\mathrm{T}}= x'_{e'}\) and \(x_{\mathrm{H}}= 
\ell_{e'}-x'_{e'}\),
then "gluing" the two ends together forming a single 
bond. The solution on $e'$ is 
then
\begin{equation} \label{psi}
    \psi_{e'}(x_{e'})=
    \begin{cases}
    a^{\mathrm{in}}_{\mathrm{T}}
    e^{- i k_+ x_{e'}}
    +\left(\sigma_{\mathrm{TH}}
    a^{\mathrm{in}}_{\mathrm{H}}+
    \sigma_{\mathrm{TT}}
    a^{\mathrm{in}}_{\mathrm{T}}
    \right)e^{i k_+ x_{e'}}
    & \text{for \(x_{e'}< x'_{e'}\);}\\
    a^{\mathrm{in}}_{\mathrm{H}}
    e^{- i k_+ (\ell_{e'}-x_{e'})}
    +\left(\sigma_{\mathrm{HH}}
    a^{\mathrm{in}}_{\mathrm{H}}+
    \sigma_{\mathrm{HT}}a^{\mathrm{in}}_{\mathrm{T}}
    \right)e^{i k_+ (\ell_{e'}-x_{e'})}
    & \text{for \(x_{e'}>x'_{e'}\).}
    \end{cases}
\end{equation}
One determines \(a^{\mathrm{in}}_{\mathrm{H}}\) and 
\(a^{\mathrm{in}}_{\mathrm{T}}\) by fulfilling equation \eqref{Definition_Green} 
at \(x_{e'}=x'_{e'}\);
this leads to the following conditions:\\ 
i. continuity at \( x_{e'}= x'_{e'}\)
\begin{equation}
    \lim_{\alpha \to 0^+}\left[\psi_{e'}(x'_{e'}+ \alpha)-
     \psi_{e'}(x'_{e'}- \alpha)\right] = 0;
\end{equation}
ii. a discontinuity of the derivatives of the form
\begin{equation}
    \lim_{\alpha\to 0 ^+} \left[
    \frac{d\psi_{e'}\left( 
        x'_{e'}+\alpha
    \right)}{dx_{e'}}-
    \frac{d\psi_{e'}\left( 
        x'_{e'}-\alpha
    \right)}{dx_{e'}}\right]=1\ .
\end{equation}
These two conditions result in a non-homogeneous system of linear
equations for the two incoming scattering amplitudes.
The unique solution of this system is
\begin{subequations}
\label{Greens_coefficients}
\begin{align}
        a^{\mathrm{in}}_{\mathrm{T}}=&
        \frac{e^{ik_+ \ell_{e'}}
        \left(e^{-ik_+(\ell_{e'}-x'_{e'})} +
        \sigma_{\mathrm{HH}} e^{ik_+(\ell_{e'}-x'_{e'})}
        -\sigma_{\mathrm{TH}}e^{ik_+x'_{e'}}
        \right)}{2ik_+
        \left[(1-e^{ik_+\ell_{e'}}
        \sigma_{\mathrm{HT}})(1-e^{ik_+\ell_{e'}} \sigma_{\mathrm{TH}})- e^{2ik_+\ell_{e'}} 
        \sigma_{\mathrm{HH}}\sigma_{\mathrm{TT}})
        \right]}
        \nonumber\\
        =
        &
        \frac{1}{2ik_+}
        \left[ 
        e^{ik_+ x'_{e'}}
        \left[ \frac{\mathbb{I}}{\mathbb{I}- \mathbf{U}(k_+)}
        \right]_{e'_- e'_-}
        +
        e^{ik_+ ( \ell_{e'}-x'_{e'})}
         \left[ \frac{\mathbb{I}}{\mathbb{I}- \mathbf{U}(k_+)}
        \right]_{e'_- e'_+}
        \right]\\
        a^{\mathrm{in}}_{\mathrm{H}}=&
        \frac{e^{ik_+ \ell_{e'}}\left(
        e^{-ik_+x'_{e'}} +
        \sigma_{\mathrm{TT}} e^{ik_+ x'_{e'}}
        -\sigma_{\mathrm{HT}}e^{ik_+(\ell_{e'}-x'_{e'})}
        \right)}{2ik_+
        \left[(1-e^{ik_+\ell_{e'}}\sigma_{\mathrm{HT}})(
        1-e^{ik_+\ell_{e'}} \sigma_{\mathrm{TH}})- e^{2ik_+\ell_{e'}} 
        \sigma_{\mathrm{HH}}\sigma_{\mathrm{TT}})
        \right]}
        \nonumber\\
        =
        &
        \frac{1}{2ik_+}
        \left[ 
        e^{ik_+(\ell_{e'} - x'_{e'})}
        \left[ \frac{\mathbb{I}}{\mathbb{I}- \mathbf{U}(k_+)}
        \right]_{e'_+ e'_+}
        +
        e^{ik_+ x'_{e'}}
         \left[ \frac{\mathbb{I}}{\mathbb{I}- \mathbf{U}(k_+)}
        \right]_{e'_+ e'_-}
        \right]\ .
    \end{align}
\end{subequations}
The derivation of the expressions involving 
\((\mathbb{I}- \mathbf{U}(k_+))^{-1}\), the resolvent matrix of the quantum map, 
can be found in \ref{appA}.
Inserting (\ref{Greens_coefficients}) into (\ref{psi}) and extending the solution 
to the entire graph using (\ref{rho_definition}), the Green's function of the 
compact graph \(\mathcal{G}\) can finally be written in the form
\begin{eqnarray}
    \label{compact_Greens_function}
    G(\mathbf{x},\mathbf{x'}, E_+)=
    \frac{1}{2k_+i}&
    \left[
     \delta_{ee'} e^{i k_+ |x_{e} -x'_{e'}|}+ e^{i k_+(x_e-x'_{e'}-
     \ell_e+\ell_{e'})}
     \left[ 
     \frac{\mathbf{U}(k_+)}{\mathbb{I}- \mathbf{U}(k_+)}
     \right]_{e_+ e'_+}\right.
     \nonumber\\
     &\;+
     e^{-i k_+(x_e-x'_{e'})}
     \left[ \frac{\mathbf{U}(k_+)}{\mathbb{I}- \mathbf{U}(k_+)}\right]_{e_- 
     e'_-}\nonumber\\
     &\;+ e^{i k_+(x_e+x'_{e'}- \ell_e)}
     \left[ \frac{\mathbf{U}(k_+)}{\mathbb{I}- \mathbf{U}(k_+)}\right]_{e_+ e'_-}
     \nonumber \\
     &\;+
     \left.
     e^{-i k_+(x_e+x'_{e'}- \ell_{e'})}
     \left[ \frac{\mathbf{U}(k_+)}{\mathbb{I}- \mathbf{U}(k_+)}\right]_{e_- e'_+}
     \right].
\end{eqnarray}
This is our main result in this section. We give here for the first time a closed 
form expression of the Green's function on a graph following the recipe from 
Barras and Gaspard \cite{BG01}.  

By formally expanding \( \frac{\mathbf{U}}{\mathbb{I}-\mathbf{U}}= 
\sum_{n=1}^\infty 
\mathbf{U}^n \), one may 
express the Green's function as a sum over paths \(p\) on the metric graph 
starting 
at \(\mathbf{x}'\) and ending at \( \mathbf{x}\), that is,
\begin{equation} \label{green_orbits}
   G(\mathbf{x},\mathbf{x'}, E_+)=\frac{1}{2k_+ i}
   \sum_{p} A_p(k_+) e^{i L_p k_+}\, .
\end{equation}
Here, \(L_p\) is the metric length of the path and 
the amplitude 
\(A_p\) 
is the product of all scattering amplitudes along the trajectory.
If \(e=e'\), the direct path between $x_{e'}$ and $x'_{e'}$ 
has \(L_p= |x_{e'}-x'_{e'}|\) and \(A_p=1\). Eq.\ (\ref{green_orbits}) is the 
starting point for the investigations in \cite{BG01}, which, however, makes it 
necessary to do an explicit summation over all possible paths - in general a 
cumbersome task.
Note also that this expansion converges only if the imaginary part of
\(k_+\) is positive and these expressions thus require a limit if used for real
wave numbers. This is all well known for similar expansions into 
sums over paths in trace formulae and scattering systems, we refer to the textbook 
\cite{BK13} and references therein.

Finally, let us shortly discuss the pole structure of the Green's function. 
For a compact graph, the eigenvalue spectrum is a discrete countable set 
\(\{E_0,E_1,\dots\}\). Let us assume that there are no 
degeneracies and all eigenvalues are positive, that is, \(E_n>0\).
The spectral decomposition of the Schr\"odinger operator
\(\hat{H}\) allows us to write the resolvent operator as
\begin{equation}
    (E_+-\hat{H})^{-1}= \sum_{n=0}^\infty \frac{\hat{P}_n}{E_+-E_n}
\end{equation}
where \(\hat{P}_n\) is the projection operator onto the subspace spanned 
by the \(n\)-th eigenvector. For the Green's function this implies
\begin{equation}
  G(\mathbf{x},\mathbf{x'}, E_+)=
  \sum_{n=0}^\infty \frac{P_n(\mathbf{x},\mathbf{x'})}{E_+-E_n} 
  \label{pole_structure}
\end{equation}
where \(P_n(\mathbf{x},\mathbf{x'})\) is the integral kernel of \(\hat{P}_n\).
Let us now show that \eqref{compact_Greens_function} and 
\eqref{pole_structure} are indeed equivalent.
We start
by considering the limit
\(E_+ \to E_n\) for some given eigenvalue \(E_n=k_n^2\) and by showing 
that the singular
part of the Green's function \eqref{compact_Greens_function}
in this limit is given by 
\(\frac{P_n(\mathbf{x},\mathbf{x'})}{E_+-E_n}\). 
Let us extract first the singular part of the matrix 
\begin{equation}
    \frac{\mathbf{U}(k_+)}{\mathbb{I}- \mathbf{U}(k_+) }\sim
    \frac{\mathbf{P}}{-i(k_+-k_n) C} \ .
    \label{singular_part}
\end{equation}
Here, \(\mathbf{P}=\mathbf{b}^{\mathrm{in}}{\mathbf{b}^{\mathrm{in}}}^\dagger\) 
is the projection matrix with matrix elements
on the
corresponding unit eigenvector \(\mathbf{U}(k_n) 
\mathbf{b}^{\mathrm{in}}=
\mathbf{b}^{\mathrm{in}}
\)
and 
\begin{equation}
 C= {\mathbf{b}^{\mathrm{in}}}^{\dagger}\left[k_n \mathbf{L}
    +  \sin(k_n \mathbf{L})\boldsymbol{\Pi} 
    \right] 
    \mathbf{b}^{\mathrm{in}}    >0
    \label{normalization_constant}
\end{equation}
is a positive constant and $\mathbf{L}$ is a $2 {N}_B$
dimensional diagonal matrices with diagonal entries $\ell_e$. 
We refer to \ref{normalization-derivation} for a detailed derivation of
\eqref{singular_part} and \eqref{normalization_constant}. 
With \(2 k_+ (k_+ -k_n) \sim E_+ -E_n\) one then finds
\begin{align}
  G(\mathbf{x},\mathbf{x'}, E_+)\sim &
  \frac{
  \big(
  a_{e_-}^{\mathrm{in}} e^{-i k_n x_e} 
  +a_{e_+}^{\mathrm{in}} e^{i k_n (x_e-\ell_e)} 
  \big)^* 
  \big( 
  a_{e'_-}^{\mathrm{in}} e^{-i k_n x_{e'}} 
  +a_{e'_+}^{\mathrm{in}} e^{i k_n (x_{e'}-\ell_{e'})} 
  \big)}{C(E_+ - E_n)}\nonumber\\
  = & \frac{P_n(\mathbf{x},\mathbf{x'})}{E_+-E_n}\ ,
\end{align}
where the last equality requires that the constant \(C\) gives the correct
normalization of the projection kernel \(P_n(\mathbf{x},\mathbf{x'})\).
This is equivalent to \(\sum_{e\in \mathcal{E}} \int_0^{\ell_e} 
P_n((e,x_e),(e,x_e)) dx_e =1\) 
which is easily checked by direct calculation.
Repeating this calculation for \(E_+\) near to all other energy eigenvalues
shows that expressions \eqref{compact_Greens_function} and \eqref{pole_structure}
have the same poles and the same residues. 
Both expressions can be continued analytically to the lower half plane where
the imaginary part of the energy is negative. 
They are thus 
equivalent up to an entire function \( F(E)\), (i.e., it is analytic in the whole complex plane).
As both \eqref{compact_Greens_function} and \eqref{pole_structure}
vanish in the limit \(E_i \to \pm \infty\), the same must be true for their difference 
\(F(E)\). The entire function that vanishes in these limits for all 
\(E_r\)  is \(F(E)=0\).

\subsection{Construction of the Green's function for open scattering graphs}

The construction of the Green's function on an open scattering graph
follows analogously.
In this case, our assumption that the energy has a positive imaginary part 
together 
with the requirement of square integrability leads to outgoing boundary
conditions along the leads. That is, the amplitudes of incoming plane waves need 
to 
vanish, as these would lead to exponentially increasing contributions.
These conditions are 
straight forward to implement and we can go through the same construction 
as for the 
compact graph. 
A short-cut is obtained by first replacing
each lead \(e \in \mathcal{L}\) by an edge of finite length with a dangling vertex 
of degree 
one and choosing some self-adjoint boundary conditions at the 
dangling vertices. 
This results in an auxiliary compact quantum graph as described in 
the previous section.
The Green's function of the auxiliary quantum graph is then given by 
\eqref{compact_Greens_function}. Clearly, the solution depends on the 
lengths that have been introduced for the leads as parameters. 
Next, one sends the introduced edge lengths to infinity. 
Because the imaginary part of the wave number 
is positive \(\mathrm{Im}\ k_+ >0\)  the  corresponding phase factors then decay as
\( e^{i k_+ \ell_e} \to 0\) as \(\ell_e \to \infty\). 
In this limit any dependence on the arbitrary choice of boundary conditions at the 
dangling vertices disappears and
what remains is the Green's function of the open graph.  
We refer to \ref{open_GF_details} for the details of the calculation
which results in
\begin{multline}
    \label{open_Greens_function}
    G(\mathbf{x},\mathbf{x'}, E_+)= \frac{1}{2k_+ i} \times \\
    \begin{cases}
    \delta_{e,e'}\, e^{i k_+ |x_{e} -x'_{e'}|}+
     e^{i k_+(x_{e}+x'_{e'})}
     \left[ 
      \mathbf{U}(k_+)_{\mathcal{LL}} +
      \mathbf{U}(k_+)_{\mathcal{LB}}
      \frac{\mathbb{I}}{\mathbb{I}- \mathbf{U}(k_+)_{\mathcal{BB}}}
      \mathbf{U}(k_+)_{\mathcal{BL}}
     \right]_{e e'}
     \\
    \qquad\hfill \text{if \(e, e' \in \mathcal{L}\),\qquad\; \quad}\\[0.5cm]
     e^{i k_+(x_e-x'_{e'}+\ell_{e'})}
     \left[\mathbf{U}(k_+)_{\mathcal{LB}} 
     \frac{\mathbb{I}}{\mathbb{I}- \mathbf{U}(k_+)_{\mathcal{BB}}}\right]_{e e'_+}+
     e^{i k_+(x_e+x'_{e'})}
     \left[ 
     \mathbf{U}(k_+)_{\mathcal{LB}} 
     \frac{\mathbb{I}}{\mathbb{I}- \mathbf{U}(k_+)_{\mathcal{BB}}}\right]_{e e'_-}
     \\
     \qquad \hfill \text{if \(e \in \mathcal{L}\) 
     and \(e' \in \mathcal{B}\),}\\[0.5cm]
      e^{-i k_+(x_e-x'_{e'})}
     \left[ \frac{\mathbb{I}}{\mathbb{I}- \mathbf{U}(k_+)_{\mathcal{BB}}}
      \mathbf{U}(k_+)_{\mathcal{BL}}
     \right]_{e_- e'} +
     e^{i k_+(x_e+x'_{e'}- \ell_e)}
     \left[ \frac{\mathbb{I}}{\mathbb{I}- \mathbf{U}(k_+)_{\mathcal{BB}}}
      \mathbf{U}(k_+)_{\mathcal{BL}}
     \right]_{e_+ e'}
     \\
     \qquad \hfill \text{if \(e \in \mathcal{B}\) 
     and \(e' \in \mathcal{L}\),}\\[0.5cm]
     \delta_{ee'}\, e^{i k_+ |x_{e} -x'_{e'}|} +
     e^{i k_+(x_e-x'_{e'}-\ell_e+\ell_{e'})}
     \left[ \frac{\mathbf{U}(k_+)_{\mathcal{BB}}}{\mathbb{I}- 
     \mathbf{U}(k_+)_{\mathcal{BB}}}\right]_{e_+ e'_+}+
      e^{-i k_+(x_e-x'_{e'})}
     \left[ \frac{\mathbf{U}(k_+)_{\mathcal{BB}}}{\mathbb{I}- 
     \mathbf{U}(k_+)_{\mathcal{BB}}}\right]_{e_- e'_-}\\[.1cm]
     \
     + \, e^{i k_+(x_e+x'_{e'}- \ell_e)}
     \left[ \frac{\mathbf{U}(k_+)_{\mathcal{BB}}}{\mathbb{I}- 
     \mathbf{U}(k_+)_{\mathcal{BB}}}\right]_{e_+ e'_-}+
     e^{-i k_+(x_e+x'_{e'}- \ell_{e'})}
     \left[ \frac{\mathbf{U}(k_+)_{\mathcal{BB}}}{\mathbb{I}- 
     \mathbf{U}(k_+)_{\mathcal{BB}}}\right]_{e_- e'_+}
    \\
    \qquad \hfill \text{if  \(e,e' \in \mathcal{B}\).\qquad\; \quad}
    \end{cases}
\end{multline}
If the energy spectrum of the graph is continuous these expressions are regular and
the limit \(\mathrm{Im}\ k_+ \to 0^+\) can be performed by just choosing
\(k_+\to k \in \mathbb{R}\). A similar expression for energies \(E_-^2 = k_-^2\)
with negative imaginary parts
may be obtained in the same way. More directly, it can be obtained from the symmetry
\eqref{GF_symmetry}.
Note that it will have a different limit as \(k_-\) approaches
the real axis.\\ 
The energy spectrum of an open graph may contain a discrete set \(\{ E_0,E_1,\dots\}\) 
of bound states in the continuum.
These have square integrable eigenfunctions and they thus vanish on the leads.
The Green's function for \(E\) close to any of these energy eigenvalues 
will have poles just as in the compact case that we discussed in the previous section.
And the calculation there applies here as well.
If either \(\mathbf{x}\) or \(\mathbf{x'}\) is chosen on a lead the expression for
the Green's function should remain regular as \(E\to E_n=k_n^2\) which is not obvious
from the given explicit expressions above which contain the inverse 
\(\left(\mathbb{I} - U(k_+)_{\mathcal{BB}}\right)^{-1}\).
We will show regularity explicitly if both \(\mathbf{x}\) and \(\mathbf{x'}\)
are on the leads. In that case the expression above reduces to
\begin{equation}
      G(\mathbf{x},\mathbf{x'}, E_+)= \frac{1}{2k_+ i} \left[
    \delta_{e,e'}\, e^{i k_+ |x_{e} -x'_{e'}|}+
     e^{i k_+(x_{e}+x'_{e'})}
     \boldsymbol{\sigma}(k_+)_{e,e'}
     \right]\ .
     \label{GF_leads}
\end{equation}
We will show in the following section that the scattering matrix is indeed
regular as \(k\to k_n\) for at a bound state.
Regularity in the case that one point is on a lead and the other on a bond can be shown as 
well using essentially the same tools but we will leave this to the reader.

\section{Regularisation schemes for perfect scars} \label{sec: bound states}

\subsection{Bound states in the continuum}
The eigenstates of a quantum graph are generally supported 
on all edges of a graph as long as the graph is fully connected. 
However, it is not too difficult to construct graphs which 
have eigenstates that are non-zero 
exclusively on a compact subgraph \(\mathcal{S}\), but vanish 
exactly on the rest \(\mathcal{R}\) of the edges. 
We call such an eigenstate a {\em perfect scar} of the graph. 
These states exist, for example, on quantum graphs with Kirchhoff-Neumann 
conditions where the subgraph \(\mathcal{S}\) is a cycle on which all edge lengths 
are rationally dependent. In that case, the cycle edge lengths are an integer 
multiple of a minimal length \(\ell_e = n_e \ell_0\). At wave number 
\(\tilde{k} = 2\pi/\ell_0\) 
(or any integer multiple of it), one may then set
\begin{equation}
    \psi_e(x_e) =
    \begin{cases}
    \pm \sin(\tilde{k} x_e) & \text{if \(e\) belongs to the cycle of 
    \(\mathcal{S}\);}\\
    0 & \text{if e belongs to \(\mathcal{R}\)}.
    \end{cases}
\end{equation} 
Here the signs $\pm$ can be chosen to satisfy the flux conservation condition.

Since the union of \(\mathcal{S}\) and \(\mathcal{R}\) make up the total graph 
\(\mathcal{G}\), it is natural to express the quantum map in the block-form
\begin{equation}
    \mathbf{U}(k)= 
    \begin{pmatrix}
       \mathbf{U}(k)_{\mathcal{RR}} & \mathbf{U}(k)_{\mathcal{RS}}\\
       \mathbf{U}(k)_{\mathcal{SR}} & \mathbf{U}(k)_{\mathcal{SS}}
    \end{pmatrix}
\end{equation}
with appropriate permutations applied.
In general there is perfect scar on the subgraph \(\mathcal{S}\) at energy 
\(E=k^2>0\), 
if the block \(\mathbf{U}(k)_{\mathcal{SS}}\) 
has an eigenvector
\(\mathbf{a}_{\mathcal{S}}^{\mathrm{in}}\) with unit eigenvalue 
\( \mathbf{U}(k)_{\mathcal{SS}} \mathbf{a}_S^{\mathrm{in}} = 
\mathbf{a}_{\mathcal{S}}^{\mathrm{in}}\). 
The unitarity
of the full quantum map then implies that 
\(\mathbf{U}(k)_{\mathcal{RS}} \mathbf{a}_{\mathcal{S}}^{\mathrm{in}}=0\) vanishes.
One may extend $\mathbf{a}_{\mathcal{S}}^{\mathrm{in}}$
to an eigenvector of the full map by setting 
\(\mathbf{a}_{\mathcal{R}}^{\mathrm{in}}=0\) resulting in the vanishing of wave 
amplitudes on edges that do not belong to \(\mathcal{S}\). 

For open graphs, a perfect scar at a wavenumber \(k_0>0\) is a bound state in the 
continuum and this situation is again straight forward to construct, such as by 
using 
the cycle example above.
In this case, one may take \(\mathcal{R}\) to contain all leads
and \(\mathcal{S}\) to be a sub-graph containing a sub-set of the finite bonds.

Throughout the previous sections, we assumed that the matrix 
\(\mathbb{I}-\mathbf{U}(k)_{\mathcal{BB}}\) is invertible, which 
it is generically the case as \(\mathbf{U}(k)_{\mathcal{BB}}\) is a block of 
a unitary matrix.  
However, a perfect scar exists, if and only if 
\(\mathbf{U}(k)_{\mathcal{BB}}\) 
has an eigenvalue one at the wave number \(k=k_0\). Even in the case of 
``almost'' perfect scars (with small nonzero entries for 
\(\mathbf{a}_{\mathcal{R}}^{\mathrm{in}}\)), matrix inversion
may cause large numerical errors when inverting 
\(\mathbb{I}-\mathbf{U}(k)_{\mathcal{BB}}\). To deal with this issue, 
we describe a regularisation scheme of the scattering matrix in the following 
section. This is important when dealing with open quantum graphs and when 
constructing
Green's function both in the compact and open case. The approach may also be used
to find the regular part of the Green's function in compact quantum graphs when
the energy is in the eigenvalue spectrum. (By regular part, we refer to the Green's 
function where the contribution from the pole at the energy has been removed).
We will focus on the regularization of the scattering matrix, as the other 
applications can all be derived from there when needed.

\subsection{Regularization of the scattering approach at a bound state} 
\label{regularisation}

We will show in this section that scattering solutions of the form 
(\ref{scattering_amplitudes_leads})
are well defined at $k=k_0$ even in the presence of a 
bound state at that wave number. We show in \ref{AppB} that the scattering 
matrix can be regularised across a whole $k$ interval containing $k_0$.  

Consider a non-degenerate bound state at wave number $k=k_0$ with wave 
amplitudes \( \mathbf{b}_{\mathcal{B}}^{\mathrm{in}}\)
such that,
\begin{equation}
    \mathbf{U}(k_0)_{\mathcal{BB}}\, \mathbf{b}_{\mathcal{B}}^{\mathrm{in}}=
    \mathbf{b}_{\mathcal{B}}^{\mathrm{in}}\ .
\end{equation}
As discussed in the previous section, the unitarity of the quantum map 
\(\mathbf{U}(k)\) implies
\begin{equation} \label{orthogonal}
    \mathbf{U}(k_0)_{\mathcal{LB}}
    \mathbf{b}_{\mathcal{B}}^{\mathrm{in}}=0 
    \qquad \text{and}
    \qquad 
    {\mathbf{b}_{\mathcal{B}}^{\mathrm{in}}}^{\dagger}
    \mathbf{U}(k_0)_{\mathcal{BL}}=0 , 
\end{equation}
that is, incoming waves 
$\mathbf{a}^{\mathrm{in}}_{\mathcal{L}}$ in the leads can not couple into the 
bound state $\mathbf{b}_{\mathcal{B}}^{\mathrm{in}}$ and the bound state can 
not couple back out. 
Let us assume for simplicity that the perfect scar described by
\(\mathbf{b}_{\mathcal{B}}^{\mathrm{in}}\) is not degenerate 
and introduce the 
idempotent, Hermitian \(2N_{\mathcal{B}} 
\times 2N_{\mathcal{B}}\) projection matrix 
\begin{equation} \label{projectionP}
    \mathbf{P}\equiv \mathbf{b}_{\mathcal{B}}^{\mathrm{in}}
    {\mathbf{b}_{\mathcal{B}}^{\mathrm{in}}}^{\dagger}
\end{equation}
and its orthogonal complement
\begin{equation}\label{projectionQ}
    \mathbf{Q}=\mathbb{I}-\mathbf{P}\ .
\end{equation}
The methods below can be generalised to situations where more than one perfect 
scar exists
at the same wave number \(k_0\), such as, if all edge lengths are rationally 
related 
in a large graph with Neumann-Kirchhoff matching conditions. 
Writing Eq.\ (\ref{rho_definition}) in the form
\begin{equation}\label{internal}
 \left(\mathbb{I}-\mathbf{U}(k)_{\mathcal{BB}}\right) 
 \mathbf{a}^{\mathrm{in}}_{\mathcal{B}} = 
   \mathbf{U}(k)_{\mathcal{BL}} \mathbf{a}^{\mathrm{in}}_{\mathcal{L}},
\end{equation}
we find that the solution $\mathbf{a}^{\mathrm{in}}_{\mathcal{B}}$ is not unique 
at $k=k_0$ as both
\begin{equation} 
\mathbf{P} \left(\mathbb{I}-\mathbf{U}(k_0)_{\mathcal{BB}}\right) =0 \quad 
\mbox{and} \quad \mathbf{P}  \mathbf{U}(k_0)_{\mathcal{BL}} = 0,
\end{equation}
which follows directly from (\ref{orthogonal}). This implies, that 
for any solution $\mathbf{a}^{\mathrm{in}}_{\mathcal{B}}$ 
of Eq.\ (\ref{internal}), $\mathbf{a}^{\mathrm{in}}_{\mathcal{B}} + 
\alpha \mathbf{b}_{\mathcal{B}}^{\mathrm{in}}$, $\alpha \in \mathbb{C}$, is also a 
solution. 
However, a unique solution $\tilde{\mathbf{a}}^{\mathrm{in}}_{\mathcal{B}}$ exists 
for the reduced system of equations
\begin{equation} \label{YQ}
  \mathbf{Y}_Q(k_0)\, \tilde{\mathbf{a}}^{\mathrm{in}}_{\mathcal{B}} = 
   \mathbf{U}(k_0)_{\mathcal{BL}}\, \mathbf{a}^{\mathrm{in}}_{\mathcal{L}}
\quad \mbox{with} \quad \mathbf{Y}_Q(k_0) = \mathbf{Q}\left(\mathbb{I}-\mathbf{U}
(k_0)_{\mathcal{BB}}\right)\mathbf{Q}.
\end{equation}
As \(\mathbf{Y}_Q(k_0)\mathbf{b}_{\mathcal{B}}^{\mathrm{in}}=0\),
its standard inverse does not exist. One may invert it in the subspace
orthogonal to \(
\mathbf{b}_{\mathcal{B}}^{\mathrm{in}}\).
Let us define (with mild abuse of notation)
\begin{equation}
   \mathbf{Y}_Q(k_0)^{-1}=
   \mathbf{Q}\frac{\mathbb{I}}{
   \mathbb{I}-\mathbf{QU}(k_0) \mathbf{Q} }
   \mathbf{Q}
   \label{YQ-inverse-definition}
\end{equation}
as the unique \(2N_{\mathcal{B}} 
\times 2N_{\mathcal{B}}\) matrix with
by \( \mathbf{Y}_Q(k_0)^{-1} \mathbf{Y}_Q(k_0)= \mathbf{Q} =
\mathbf{Y}_Q(k_0) \mathbf{Y}_Q(k_0)^{-1}   \) and
\( \mathbf{Y}_Q(k_0)^{-1} \mathbf{P}= 0 =
\mathbf{P} \mathbf{Y}_Q(k_0)^{-1}   \). 
As 
$\mathbf{U}(k_0)_{\mathcal{LB}}\mathbf{P} = 0$, one obtains a well-defined 
scattering solution for Eq.\ (\ref{scattering_amplitudes_leads}), that is,
\begin{equation}
    \mathbf{a}(k)^{\mathrm{out}}_{\mathcal{L}} = \mathbf{U}(k_0)_{\mathcal{LB}}\, 
    \tilde{\mathbf{a}}^{\mathrm{in}}_{\mathcal{B}} .
\end{equation}
We may thus write the scattering matrix (\ref{scattering_matrix}) in the form
\begin{equation} \label{s-matrix-reg}
    \boldsymbol{\sigma}(k_0) =
    \mathbf{U}(k_0)_{\mathcal{LL}}
    +\mathbf{U}(k_0)_{\mathcal{LB}} \mathbf{Y}_Q(k_0)^{-1}
    \mathbf{U}(k_0)_{\mathcal{BL}}\ .
\end{equation}
For an in-depth discussion of the regularity of the scattering matrix as 
$k\to k_0$, see \ref{AppB}.

\section{Worked examples}
\label{sec:examples}
In this section we explicitly construct the scattering matrices of two open 
quantum graphs which 
contain perfect scars.
Expressions for the Green's function on the leads 
follow directly using \eqref{GF_leads}.

\subsection{Open lasso}
\begin{figure}[H]
\centering
\includegraphics[width = 0.4\textwidth]{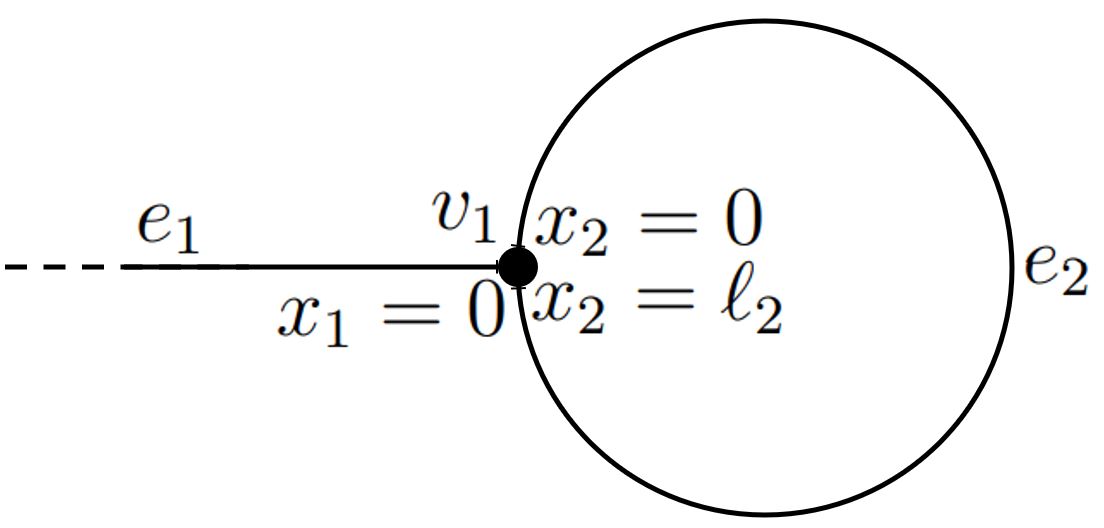}
\caption{An open lasso graph constructed from two edges $e_{1}$ and $e_{2}$
where $e_{1}$ is a lead and $e_{2}$ is an bond. Both edges are connected to the 
same vertex $v_{1}$ where edge $e_{2}$ has both ends connected forming a loop 
wherein bound states can exist in the continuum.
\label{fig:Lasso Graph}}
\end{figure} Consider the open lasso quantum graph illustrated in figure 
\ref{fig:Lasso Graph}. The coordinate \(x_1\ge 0\) 
runs along the lead with \(x_1=0\) at the vertex \(v_1\) and the coordinate 
\(x_2\in [0,\ell_2]\) runs along the loop such that 
\(x_2=0\) and \(x_2=\ell_2\) are the endpoints at the vertex \(v_1\).
At the vertex, we enforce Neumann boundary conditions, as expressed in 
(\ref{Kirchoff-Neumann BC}), leading to the quantum map written in block form as
\begin{equation} \label{ulasso}
\mathbf{U}(k)= 
\begin{pmatrix}
    \begin{array}{c|cc}
  -\frac{1}{3} &  \frac{2}{3} & \frac{2}{3}\\[.1cm]
  \hline
  \rule{0cm}{.55cm}\frac{2 e^{ik\ell_2}}{3} &\frac{2 e^{ik\ell_2}}{3} &
  -\frac{e^{ik\ell_2}}{3}\\[.1cm]
  \frac{2 e^{ik\ell_2}}{3}  &-\frac{e^{ik\ell_2}}{3}&
  \frac{2 e^{ik\ell_2}}{3}
    \end{array}
\end{pmatrix}
\equiv 
\begin{pmatrix}
    \mathbf{U}_{\mathcal{LL}} & \mathbf{U}_{\mathcal{LB}}\\
    \mathbf{U}(k)_{\mathcal{BL}}& \mathbf{U}(k)_{\mathcal{BB}}\\
\end{pmatrix}.
\end{equation}
In the construction of the scattering matrix and the Green's function, 
one needs to invert the matrix 
\( \mathbb{I} - \mathbf{U}(k)_{\mathcal{BB}}\) which yields
\begin{equation} \label{1-ulasso}
\frac{\mathbb{I}}{\mathbb{I} - \mathbf{U}(k)_{\mathcal{BB}}}=
\begin{pmatrix}
    \frac{3-2 e^{ik\ell_2}}{\left(e^{ik\ell_2}-1\right)\left(e^{ik\ell_2}-3\right)}
    &-\frac{e^{ik\ell_2}}{\left(e^{ik\ell_2}-1\right)\left(e^{ik\ell_2}-3\right)}
    \\[.3cm]
    -\frac{e^{ik\ell_2}}{\left(e^{ik\ell_2}-1\right)\left(e^{ik\ell_2}-3\right)}
    &
    \frac{3-2 e^{ik\ell_2}}{\left(e^{ik\ell_2}-1\right)\left(e^{ik\ell_2}-3\right)}
\end{pmatrix}
\end{equation}
and is well defined as long as \( e^{i k\ell_2} \neq 1\), that is, 
if \(k \ne k_n = 2\pi n/ \ell_2\) for
\(n=1,2,\dots\). The reason for this
is the existence of perfect scars on the loop which here lead to bound states in 
the continuum of scattering states.
These bound state wave functions are given as
\begin{subequations}
\begin{align}
\psi_{e_1}(x_1)=& 0,\\
\psi_{e_2}(x_2)=& \sqrt{\frac{2}{\ell_2}}\sin(k_n x_2)\ .
\end{align}
\end{subequations}
The continuum of scattering states exists for all wave numbers \(k>0\)
and is given by
\begin{subequations}
\begin{align}
\psi_{e_1}(x_1)=& e^{-ikx_1}+ \boldsymbol{\sigma}(k) e^{ik x_1},\\
\psi_{e_2}(x_2)=& \boldsymbol{\rho}(k)_{2_{+} 1} e^{ik (x_2-\ell_2)}+
\boldsymbol{\rho}(k)_{2_{-} 1}e^{-ik x_2}
\ .
\end{align}
\end{subequations}
where
\begin{equation}
    \boldsymbol{\rho}(k)=
    \frac{\mathbb{I}}{\mathbb{I} - \mathbf{U}(k)_{\mathcal{BB}}} 
    \mathbf{U}(k)_{\mathcal{BL}}
    = \begin{pmatrix}
    \frac{2 e^{ik \ell_2}}{3-e^{ik\ell_2}}\\[.1cm]
    \frac{2 e^{ik \ell_2}}{3-e^{ik\ell_2}}
    \end{pmatrix}
\end{equation}
and
\begin{equation}
 \boldsymbol{\sigma}(k)  = \mathbf{U}_{\mathcal{LL}}+ \mathbf{U}_{\mathcal{LB}} 
 \boldsymbol{\rho}(k)=
 \frac{3e^{ik\ell_2}-1}{3-e^{ik\ell_2}}\ .
\end{equation}
While the matrix
\(\frac{\mathbb{I}}{\mathbb{I} - \mathbf{U}(k)_{\mathcal{BB}}}\) is used
to find \(\boldsymbol{\rho}(k)\) and \(\boldsymbol{\sigma}(k)\) in the scattering 
approach
the poles at \(k=k_n\) have disappeared in the final results.
Note that bound states and scattering states are trivially 
orthogonal due to their symmetry under \(x_2 \mapsto \ell_2-x_2\)
(which can be viewed as a mirror symmetry of the lasso).
The bound states are odd under this symmetry as
\(\psi_1(x_1)=0\) and
\(\psi_2(x_2)=-\psi_2(\ell_2 -x_2)\) at wave numbers \(k_n\).
The scattering states are even under this symmetry for
all wave numbers \(k>0\) as
\begin{equation}
\psi_{e_2}(x_2)= \frac{4 e^{ik \ell_2/2}}{3-e^{ik\ell_2}}
\cos\left(k\frac{2x_2-\ell_2}{2}\right)=
\psi_2(\ell_2-x_2) .
\end{equation}
For completeness, we give the full Green's function for this example below, where $x_e$ (or $x'_{e'}$) are either on the lead ($e = e_1$) or on the loop ($e=e_2$). Following on from the last line in \eqref{open_Greens_function}, one obtains, using the expressions in (\ref{ulasso}) and (\ref{1-ulasso}),
\begin{multline}
    \label{Lasso Green's function}
    G_{\mathrm{lasso}}(\mathbf{x},\mathbf{x'}, E_+)= \frac{1}{2k_+ i} \times \\
    \begin{cases}
    e^{i k_+ |x_{e} -x'_{e'}|}+
     e^{i k_+(x_{e}+x'_{e'})}
     \frac{3e^{ik_{+}\ell_{2}} - 1}{3 - e^{ik_{+}\ell_{2}}}
     & \text{if \(e = e_{1}\) 
     and \(e' = e_{1}\),}\\[0.5cm]
     \frac{2}{3 - e^{ik_{+}\ell_{2}}}e^{ik_{+}x_{e_{1}}}\left(e^{ik_{+}x'_{e_{2}}} +  e^{-ik_{+}(x'_{e_{2}} - \ell_{2})}\right)
     & \text{if \(e = e_{1}\) 
     and \(e' = e_{2}\),}\\[0.5cm]
     \frac{2}{3 - e^{ik_{+}\ell_{2}}}e^{ik_{+}x'_{e_{1}}}\left(e^{-ik_{+}x_{e_{2}}} +  e^{ik_{+}(x_{e_{2}} - \ell_{2})}\right)
     & \text{if \(e = e_{2}\) 
     and \(e' = e_{1}\),}\\[0.5cm]
     \,e^{ik_+|x_{e_2} - x'_{e_2}|} + 
    \frac{2e^{ik_+\ell_{2}}}{(e^{ik_+\ell_{2}}-1)(e^{ik_+\ell_{2}} - 3)}\left[(2 - e^{ik_+\ell_{2}})\cos(k_+ (x_{e_2} - x'_{e_2}))
    \right.\\
    \qquad
    \left.-\cos(k_+(x_{e_2} + x'_{e_2} - \ell_{2}))\right]
     & \text{if \(e = e_{2}\) 
     and \(e' = e_{2}\).}\\[0.5cm]
    \end{cases}
\end{multline}

\subsection{Scattering states for an open 3-star with one lead}
\begin{figure}[H]
\centering
\includegraphics[width = 0.35\textwidth]{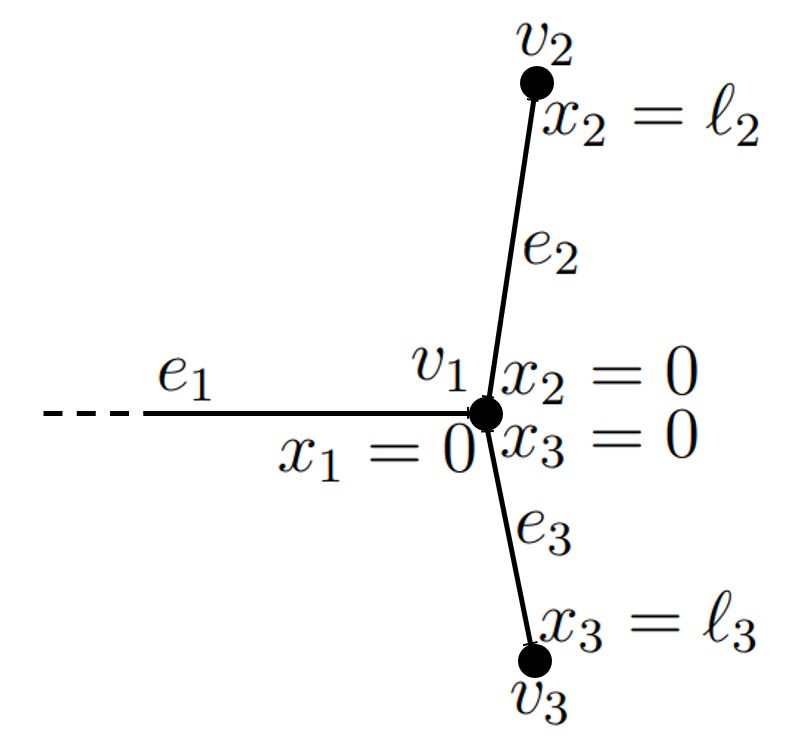}
\caption{A 3-star with one lead consists of a central
vertex \(v_1\) with three edges \(e_{n}, n=1,2,3\), attached. 
Here, \(e_1\) is a lead
and the other two edges \(e_2\) and \(e_3\) are bonds of lengths 
\(\ell_2\) and \(\ell_3\) ending in vertices \(v_2\) and \(v_3\).
\label{fig:T-Junction Graph}}
\end{figure} 
Consider the open T-junction quantum graph as illustrated in 
Figure~\ref{fig:T-Junction Graph}.
We choose the three coordinates such that
\(x_n=0\) for \(n=1,2,3\) at the central vertex \(v_1\) with
\(x_{n}=\ell_n\) at vertices \(v_{n}, n=2,3\).
We enforce Kirchhoff-Neumann boundary conditions at the central vertex as 
expressed 
in (\ref{Kirchoff-Neumann BC}) and Dirichlet boundary conditions at \(v_2, v_3\), 
that is,
\(\Sigma^{(v_n)} = -1, n=2,3\), leading to the quantum map
\begin{equation}
\mathbf{U}(k)=
\begin{pmatrix}
\begin{array}{c|cccc}
-\frac{1}{3} & 0 & 0 & \frac{2}{3} & \frac{2}{3}\\[.1cm]
    \hline \rule{0cm}{0.5cm}
    \frac{2e^{ik\ell_2}}{3}& 0&0 &- 
    \frac{e^{ik\ell_2}}{3}& \frac{2e^{ik\ell_2}}{3}\\
    \frac{2e^{ik\ell_3}}{3}& 0&0& \frac{2e^{ik\ell_3}}{3} &
    -\frac{e^{ik\ell_3}}{3}\\
    0& -e^{ik\ell_2} & 0 & 0 & 0\\
    0&  0 & -e^{ik\ell_3} & 0 &0
    \end{array}
\end{pmatrix}
\equiv 
\begin{pmatrix}
    \mathbf{U}_{\mathcal{LL}} & \mathbf{U}_{\mathcal{LB}}\\
    \mathbf{U}(k)_{\mathcal{BL}} & \mathbf{U}(k)_{\mathcal{BB}}\\
\end{pmatrix}.
\end{equation}
Computing the scattering matrix and Green's function in the scattering 
approach require that
one inverts the matrix \( \mathbb{I} - \mathbf{U}(k)_{\mathcal{BB}}\) 
which is given as
\begin{equation}
\begin{split}
& \frac{\mathbb{I}}{\mathbb{I} - \mathbf{U}(k)_{\mathcal{BB}}}=
\\
& \frac{1}{D} 
\begin{pmatrix}
    3-e^{2ik\ell_3} & 
    -2e^{ik(\ell_2+\ell_3)}
    & -(1+e^{2ik\ell_3})e^{ik \ell_2}& 2e^{ik\ell_2}\\
    -2e^{ik(\ell_2+\ell_3)}& 
    3-e^{2ik\ell_2}&2 e^{ ik \ell_3} &    -(1+e^{2ik\ell_2})e^{ik \ell_3} \\
    -(3-e^{2ik\ell_3})e^{ik\ell_2} &
    2e^{ik(2\ell_2+\ell_3)}
    & 3-e^{2ik\ell_3} & 
    -2e^{2ik\ell_2}\\
    2e^{ik(\ell_2+2\ell_3)}& 
    -(3-e^{2ik\ell_2})e^{ik\ell_3}&-2e^{2ik\ell_3}& 
    3-e^{2ik\ell_2}
\end{pmatrix} 
\end{split}
\end{equation}
where
\begin{equation}
 D= 3-e^{2ik\ell_2}-e^{2ik\ell_3}-e^{2ik(\ell_2+\ell_3)}\ . 
\end{equation}
Note that for \(e^{2 i k \ell_2}=e^{2ik\ell_3} =1\), one has \(D=0\)
making the inverse not well defined. This can only happen if the bond 
lengths are rationally related, then giving rise to a set of bound 
state in the continuum that vanish on the lead and are a sinusoidal 
wave along the two bonds with a node on the 
vertex \(v_1\). 
In either case the scattering states are given by
\begin{subequations}
\begin{align}
\psi_{e_1}(x_1)=& e^{-ikx_1}+ \boldsymbol{\sigma}(k) e^{ik x_1},\\
\psi_{e_2}(x_2)=& \boldsymbol{\rho}(k)_{2_{+} 1} e^{ik (x_2-\ell_2)}+
\boldsymbol{\rho}(k)_{2_{-} 1}e^{-ik x_2}\\
\psi_{e_3}(x_3)=& \boldsymbol{\rho}(k)_{3_{+} 1} e^{ik (x_3-\ell_3)}+
\boldsymbol{\rho}(k)_{3_{-} 1}e^{-ik x_3}
\end{align}
\end{subequations}
where
\begin{equation}
    \boldsymbol{\rho}(k)=
    \frac{\mathbb{I}}{\mathbb{I}-\mathbf{U}(k)_{\mathcal{BB}}}
    \mathbf{U}(k)_{\mathcal{BL}}
    =\frac{2}{D}
    \begin{pmatrix}
        e^{ik\ell_2}\left(1- e^{2ik \ell_3}\right)\\
        e^{ik\ell_3}\left(1- e^{2ik \ell_2}\right)\\ 
        -e^{2ik\ell_2}\left(1- e^{2ik \ell_3}\right)\\
        -e^{2ik\ell_3}\left(1- e^{2ik \ell_2}\right)
    \end{pmatrix}
\end{equation}
and
\begin{equation}
  \boldsymbol{\sigma}(k)=
    \mathbf{U}_{\mathcal{LL}}+ \mathbf{U}_{\mathcal{LB}}\boldsymbol{\rho}(k)
    = \frac{D^*}{D}e^{2ik(\ell_2+\ell_3)}\ .
\end{equation}
The scattering states are then given as
\begin{subequations}
\begin{align}
\psi_{e_1}(x_1)=& e^{-ikx_1}+ \frac{D^*}{D} e^{ik (x_1+2 \ell_2+ 2 \ell_3) },\\
\psi_{e_2}(x_2)=& 
\frac{2(1-e^{2ik \ell_2})(1-e^{2ik\ell_3})}{D}\frac{\sin(k(\ell_2-x_2))}{ 
\sin(k\ell_2)},\\
\psi_{e_3}(x_3)=& 
\frac{2(1-e^{2ik \ell_2})(1-e^{2ik\ell_3})}{D}\frac{\sin(k(\ell_3-x_3))}{ 
\sin(k\ell_3)}
\ .
\end{align}
\end{subequations}
The scattering matrix is continuous due to \(1+\boldsymbol{\sigma}(k)= 
\frac{2(1-e^{2ik \ell_2})(1-e^{2ik\ell_3})}{D} \). It is straight forward to check 
that the scattering states also behave well
near \(e^{2ik \ell_2} =e^{2ik \ell_3}=  1 \). 
Given the above scattering matrix constructions, the Green's function can be 
derived analogously to the previous example from equation (\ref{open_Greens_function}).

\section{Conclusion}\label{sec:conclusion}
To conclude, we present a simple three step procedure for generating the Green's 
function on both closed and open finite quantum graphs. 
The procedure exploits the standard scattering approach wherein the infinite 
sum of trajectories between a given source point and receiver point on the 
graph involves the inverse of a block component of the matrix defining the 
graph's quantum map. Generically, this matrix is sub-unitary and its inverse 
is well defined. Using this scattering representation, a closed form expression for 
the Green's function is given here for the first time. 
We also discuss the possibility of perfect scars and bound states
in the continuum for which the existing approaches (based on sums over 
trajectories)
diverge. We show that our closed expressions can be regularized in these cases.
This regularization scheme is important also on a practical level, as
scattering matrices of generic quantum graphs with \(\mathrm{NK}\) matching 
conditions which do not have any exact bound states still have resonances. These 
can be
arbitrarily close to bound states and they can lead to large errors
in numerical investigations if not treated with care.

We restricted ourselves here to the positive energy domain, mainly to keep the 
discussion concise and relevant - generalizations to the negative energy domain
follow along the same ideas, but require extra care as scattering matrices are no 
longer unitary. A more relevant extension of our results would be to graphs
which do not have a finite number of edges (such as infinite periodic quantum 
lattices).\\

\noindent{\bf Acknowledgement}\\
SG would like to acknowledge support by the COST action CA18232. TL thanks EPSRC 
for supporting his PhD studies. 

\appendix

\section{Derivation of coefficients in the Green's function in terms of 
the resolvent matrix of the quantum map} \label{appA}

For any given edge \(e \in \mathcal{E}\), we will denote its complement as
\begin{equation}
    \mathcal{E}^e \equiv \mathcal{E}\setminus \{e\}\ .
\end{equation}
Analogously, we write \( \mathcal{B}^e= \mathcal{B}\setminus \{e\}\)
if \(e \in \mathcal{B}\) or 
 \( \mathcal{L}^e= \mathcal{L}\setminus \{e\}\)
if \(e \in \mathcal{L}\). 
For any given edge \(e\), we may now write the quantum map in block form 
(after appropriate reordering of the directed edges), that is,
\begin{equation}
    \mathbf{U}=
    \begin{pmatrix}
       \mathbf{U}_{ee} & \mathbf{U}_{e \mathcal{B}^e}\\
       \mathbf{U}_{\mathcal{B}^e e} & \mathbf{U}_{\mathcal{B}^e \mathcal{B}^e}
    \end{pmatrix}\ ,
\end{equation}
where 
\(\mathbf{U}_{ee}\), \(\mathbf{U}_{e \mathcal{B}^e}\),
\(\mathbf{U}_{\mathcal{B}^e e}\) and 
\(\mathbf{U}_{\mathcal{B}^e \mathcal{B}^e}\)
are matrices of dimension \(2 \times 2\), \(2 \times 2(N_{\mathcal{B}}-1)\), 
\( 2(N_{\mathcal{B}}-1)\times 2\)
and \( 2(N_{\mathcal{B}}-1)\times 2(N_{\mathcal{B}}-1)\), respectively.
Eliminating the $\mathbf{a}_B^{\mathrm{in}}$ components 
in (\ref{quantization}), we can write the quantization condition with the help of the unitary  \(2\times2 \)
matrix \(\mathbf{U}(k)^{\mathrm{red},e}\) defined as
\begin{equation}
   \mathbf{U}^{\mathrm{red},e}=
   \mathbf{U}_{ee} + \mathbf{U}_{e \mathcal{B}^e}\left( \mathbb{I} - \mathbf{U}_{\mathcal{B}^e \mathcal{B}^e}\right)^{-1}
   \mathbf{U}_{\mathcal{B}^e e}.
\end{equation}
We also define an alternative reduced secular function
\begin{equation}
    \xi(k)^{\mathrm{red}, e}\equiv \det \left( \mathbb{I} - 
    \mathbf{U}(k)^{\mathrm{red},e} \right),
\end{equation}
which is related to $\xi(k)$ defined in (\ref{secular})  through the identity
\begin{equation}
    \xi(k)= \xi(k)^{\mathrm{red}, e} 
    \det\left(\mathbb{I}- \mathbf{U}(k)_{\mathcal{B}^e \mathcal{B}^e}\right).
\end{equation}
The relation above is obtained using the decomposition
\begin{equation} \label{decomp}
 \mathbb{I} - {\bf U}=
    \begin{pmatrix}
       \mathbb{I} - {\bf U}^{\mathrm{red,e}} & -{\bf U}_{e \mathcal{B}^e}\left( \mathbb{I} - {\bf U}_{\mathcal{B}^e \mathcal{B}^e}\right)^{-1}\\
       0 &\mathbb{I} 
    \end{pmatrix}
    \begin{pmatrix}
        \mathbb{I} &0 \\
        - {\bf U}_{\mathcal{B}^e e} & \mathbb{I} - {\bf U}_{\mathcal{B}^e \mathcal{B}^e}
    \end{pmatrix}.
\end{equation}
Note that the reduced quantum map ${\bf U}^{\mathrm{red},e}$ is related to the quantum scattering matrix
$\boldsymbol{\sigma}(k)$ introduced in Eq.\ (\ref{qscatter}) by 
\begin{equation} \label{uredsigma}
 {\bf U}^{\mathrm{red,e}} = \begin{pmatrix}
      {\bf U}^{\mathrm{red,e}}_{e^+e^+}&{\bf U}^{\mathrm{red,e}}_{e^+e^-}\\[.1cm]
      {\bf U}^{\mathrm{red,e}}_{e^-e^+}&{\bf U}^{\mathrm{red,e}}_{e^-e^-}
    \end{pmatrix}
    = e^{ik\ell_e}
    \begin{pmatrix}
        \sigma_{\mathrm{TH}}& \sigma_{\mathrm{TT}} \\
       \sigma_{\mathrm{HH}}& \sigma_{\mathrm{HT}}
    \end{pmatrix}.
\end{equation}

In order to obtain the second line in (\ref{Greens_coefficients}), we note
that the denominator in these expressions can be written in terms of the reduced secular
function of the compact graph, that is,
\begin{equation}
    \left[(1-e^{ik\ell_{e'}}\sigma_{\mathrm{HT}})(
    1-e^{ik\ell_{e'}} \sigma_{\mathrm{TH}})- e^{2ik\ell_{e'}} 
        \sigma_{\mathrm{HH}}\sigma_{\mathrm{TT}}\right]
        = \xi(k)^{\mathrm{red},e'}\ ,
\end{equation}
where we use the $e'$ notation as in Sec.\ \ref{sec:green_compact}.

By writing out the resolvent of the reduced $2\times 2$ quantum map, that is,
\begin{equation}
    \frac{\mathbb{I}}{\mathbb{I} - {\bf U}^{\mathrm{red},e'}}
    \equiv
    \begin{pmatrix}
        1-{\bf U}^{\mathrm{red}, e'}_{e'_+ e'_+}&
        -{\bf U}^{\mathrm{red}, e'}_{e'_+ e'_-}\\
        -{\bf U}^{\mathrm{red}, e'}_{e'_- e'_+}&
        1-{\bf U}^{\mathrm{red}, e'}_{e'_- e'_-}
    \end{pmatrix}^{-1}
    =
    \frac{1}{\xi^{\mathrm{red},e'}}
    \begin{pmatrix}
        1-{\bf U}^{\mathrm{red}, e'}_{e'_- e'_-}
        &
        {\bf U}^{\mathrm{red}, e'}_{e'_+ e'_-}\\
        {\bf U}^{\mathrm{red}, e'}_{e'_- e'_+}&
         1-{\bf U}^{\mathrm{red}, e'}_{e'_+ e'_+}
    \end{pmatrix}\, ,
\end{equation}
we can relate the terms in (\ref{Greens_coefficients}) to matrix elements of the 
inverse of the reduced quantum map using again (\ref{uredsigma}).
The expressions as given in Eq.\ (\ref{Greens_coefficients}) are now obtained 
observing in addition
\begin{equation}
    \frac{\mathbb{I}}{\mathbb{I} - {\mathbf 
    {U}}^{\mathrm{red},e'}}=
    \left[ 
    \frac{\mathbb{I}}{\mathbb{I} - {\bf U}}
    \right]_{e'e'}\, ,
\end{equation} 
which follows, for example, from the decomposition 
(\ref{decomp}).

\section{Details on the pole contribution to the Green's function in compact graphs}
\label{normalization-derivation}

In this appendix, we want to give a detailed derivation of 
equations \eqref{singular_part} and 
\eqref{normalization_constant}
that define the pole contribution of the Green's function at an energy eigenvalue
\(E_n=k_n^2\). 
With the orthogonal projector \(\mathbf{Q}=\mathbb{I}-
\mathbf{P}\) let us start by writing
\begin{align} \label{expansion}
 \frac{\mathbf{U}(k_+)}{\mathbb{I} - \mathbf{U}(k_+)}= &
 - \mathbb{I}+
  \frac{1}{\chi(k_+)} \mathbf{P}
  \nonumber \\
 &+ \mathbf{P}  
 \frac{\mathbb{I}}{\mathbb{I} - \mathbf{U}(k_+)}
 \mathbf{Q}+
 \mathbf{Q}  
 \frac{\mathbb{I}}{\mathbb{I} - \mathbf{U}(k_+)}
 \mathbf{P}
 +
 \mathbf{Q}  
 \frac{\mathbb{I}}{\mathbb{I} - \mathbf{U}(k_+)}
 \mathbf{Q} 
\end{align}
where 
\begin{equation}
  \chi(k_+)= \left({\mathbf{b}^{\mathrm{in}}}^\dagger
 \frac{\mathbb{I}}{\mathbb{I} - \mathbf{U}(k_+)}
  {\mathbf{b}^{\mathrm{in}}}\right)^{-1}
\end{equation}
and we have used that
\(\mathbf{P}={\mathbf{b}^{\mathrm{in}}}
{\mathbf{b}^{\mathrm{in}}}^\dagger \)
is a rank one projector.
We will show that, as \(k_+ \to k_n\), the only singular term in (\ref{expansion}) 
is contained in \( \frac{1}{\chi(k_+)}\mathbf{P}\).
Writing 
\begin{equation}
     \frac{\mathbb{I}}{\mathbb{I} - \mathbf{U}(k_+)}
     \left(\mathbf{P}+\mathbf{Q}\right)
     \left( \mathbf{I} -\mathbf{U}(k_+)\right)
     =
     \mathbb{I}\ ,
\end{equation}
and multiplying it from left and right with either 
\(\mathbf{P}\) or \(\mathbf{Q}\) results in four equations that may be solved for 
\begin{subequations}
    \begin{align}
        \chi(k_+)=&
        {\mathbf{b}^{\mathrm{in}}}^\dagger
        \left[
        \mathbb{I} - \mathbf{U}(k_+) -
        \mathbf{U}(k_+) \mathbf{Q} 
        \frac{\mathbb{I}}{\mathbb{I}- \mathbf{QU}
        (k_+)\mathbf{Q}} \mathbf{QU}(k_+)
        \right]
        {\mathbf{b}^{\mathrm{in}}}\\
        \mathbf{P} 
        \frac{\mathbb{I}}{\mathbb{I} - \mathbf{U}(k_+)}
        \mathbf{Q}=&
        \frac{1}{\chi(k_+)}
        \mathbf{PU}(k_+) \mathbf{Q}
        \frac{\mathbb{I}}{\mathbb{I} - \mathbf{QU}(k_+) 
        \mathbf{Q}}
        \mathbf{Q}
        \\
        \mathbf{Q} 
        \frac{\mathbb{I}}{\mathbb{I} - \mathbf{U}(k_+)}
        \mathbf{P}=&
        \frac{1}{\chi(k_+)}
        \mathbf{Q}\frac{\mathbb{I}}{\mathbb{I} - 
        \mathbf{QU}(k_+) \mathbf{Q}}
        \mathbf{Q}    \mathbf{U}(k_+)\mathbf{P}     \\
        \mathbf{Q} 
        \frac{\mathbb{I}}{\mathbb{I} - \mathbf{U}(k_+)}
        \mathbf{Q}=&\mathbf{Q}+\frac{1}{\chi(k_+)}
        \mathbf{Q}\frac{\mathbb{I}}{\mathbb{I} - 
        \mathbf{QU}(k_+) \mathbf{Q}}
        \mathbf{Q}    \mathbf{U}(k_+)\mathbf{PU}(k_+) 
        \mathbf{Q}\frac{\mathbb{I}}{\mathbb{I} - 
        \mathbf{QU}(k_+) \mathbf{Q}}\mathbf{Q}
    \end{align}
\end{subequations}
using standard properties of orthogonal projectors such as
\(\mathbf{P}^2=\mathbf{P}\), \(\mathbf{Q}^2=\mathbf{Q}\),
and \(\mathbf{PQ}=\mathbf{QP}=0\).
Now let us write \(k=k_n + \delta k\) and consider 
\(\delta k \to 0\) using the Taylor expansion
\begin{equation}
    \mathbf{U}(k_n+\delta k)=
    \mathbf{U}(k_n)+ \frac{d \mathbf{U}}{dk}(k_n) \ 
    \delta k + O((\delta k)^2) \ .
    \label{Taylor}
\end{equation}
The derivative of the quantum map \(\mathbf{U}(k)\)
can be performed explicitly. 
The latter depends on the wave number 
via phases  \(e^{i k \ell_e}\) on each edge $e$, and in general 
also 
via an explicit $k$ dependence of the vertex scattering matrices.
For the vertex scattering matrices of the form 
(\ref{general_vertex_scattering}), one finds, using standard matrix 
algebra,
\begin{equation}
\label{scattering derivative}
    \frac{d}{dk}
    \boldsymbol{\Sigma}^{(v)}(k)= 
    \frac{1}{2k} \left( \mathbb{I} - \boldsymbol{\Sigma}^{(v)}(k)^2 
    \right) \ .
\end{equation}
Then the derivative of \(\mathbf{U}(k)= e^{ik \mathbf{L}} 
\boldsymbol{\Pi} \boldsymbol{\Sigma}\) gives
\begin{equation}
    \frac{d\mathbf{U}}{dk} (k)=
    i \mathbf{L}
    \mathbf{U}(k)
    +
    \frac{1}{2k}
    \left[
    e^{ik \mathbf{L}} \boldsymbol{\Pi}
     - \mathbf{U}(k)  
     e^{-ik\mathbf{L}} \boldsymbol{\Pi}
     \mathbf{U}(k)
 \right]\ .
 \label{Qmap_derivative}
\end{equation}
At this stage we may identify that the constant \(C\)
stated in \eqref{normalization_constant} is just
\begin{equation}
    C= \frac{1}{i} {\mathbf{b}^{\mathrm{in}}}^\dagger\
    \frac{d\mathbf{U}}{dk} (k_n) {\mathbf{b}^{\mathrm{in}}}\ .
\end{equation}
The expressions \eqref{Taylor}
and \eqref{Qmap_derivative} have the following implications
\begin{subequations}
    \begin{align}
        \mathbf{PU}(k+\delta k)\mathbf{Q}= &
        O(\delta k)\\
        \mathbf{QU}(k+\delta k)\mathbf{P}= &
        O(\delta k)\\
        \chi(k+\delta k)=&
        -iC \delta k + O((\delta k)^2)
    \end{align}
\end{subequations}
such that  \(\mathbf{P} 
\frac{\mathbb{I}}{\mathbb{I} - \mathbf{U}(k)}
\mathbf{Q}\), \(\mathbf{Q} 
\frac{\mathbb{I}}{\mathbb{I} - \mathbf{U}(k)}
\mathbf{P}\) and \(\mathbf{Q} 
\frac{\mathbb{I}}{\mathbb{I} - \mathbf{U}(k)}
\mathbf{Q}\) are not singular in the limit \(\delta k \to 0\)
and we are left with the singular part
\begin{equation}
    \frac{\mathbf{U}(k_n +\delta k)}{\mathbb{I} - \mathbf{U}(k_n + \delta k)}= 
 \frac{1}{-i C \delta k} \mathbf{P}+ O((\delta k)^0)
\end{equation}
which is equivalent to the Eq.~\eqref{singular_part}
we wanted to proof in this appendix.

\section{Details of the derivation of the Green's function 
in open 
scattering graphs}
\label{open_GF_details}

In this appendix, we give details how the Green's 
function 
\eqref{open_Greens_function} for an
open scattering graph \( \mathcal{G}\) can be derived from the
Green's function \eqref{compact_Greens_function} of an auxiliary 
compact graph \( \mathcal{G}_{\mathrm{aux}}\) by sending the edge 
lengths of those edges turning into leads to infinity.
The auxiliary graph \( \mathcal{G}_{\mathrm{aux}}\) is obtained 
from the open graph
\( \mathcal{G}\) by replacing each lead by an edge of finite length 
with a 
vertex of degree one at the other end. 
For simplicity, we will put 
Neumann-Kirchhoff conditions at the vertices of degree one, the final 
results will not 
depend on this choice. For the sake of this derivation, we will 
bend the use of notation and continue to refer to `leads' and `bonds' of the 
auxiliary graph. Let us also introduce the  
\(N_{\mathcal{L}}\)-dimensional diagonal matrix 
\(\mathbf{L}_{\mathcal{L}}=
\mathrm{diag( \ell_e: e \in \mathcal{L})}\)
that contains the edge lengths of the leads.
We start from the Green's function for the auxiliary graph 
\eqref{compact_Greens_function}.
It contains four matrix elements of the matrix 
\(\mathbf{R}=\frac{{\mathbf{U}}^{\mathrm{aux}}}{\mathbb{I}-
{\mathbf{U}}^{\mathrm{aux}}}\) where we denote the 
(\(2(N_\mathcal{B}+ N_{\mathcal{L}})\)-dimensional) quantum map of 
the auxiliary 
graph
by \({\mathbf{U}}^{\mathrm{aux}}\) in order to distinguish it from 
the 
(\(2N_{\mathcal{B}}+ N_{\mathcal{L}} \)-dimensional)
quantum map \(\mathbf{U}\) of the open graph.
We suppress the dependence on
\(k_+\) here, as it can be reintroduced easily at the end of the 
calculation.
The standard way to continue the calculation would be to decompose 
the
involved matrices into blocks that correspond to three sets of 
directed edges:
directed bonds \(\mathcal{B}\), outgoing leads \(\mathcal{L}_+\) 
and incoming leads \(\mathcal{L}_-\).
For the quantum map of the auxiliary graph the structure of the 
graph then implies
\begin{equation}
    {\mathbf{U}}^{\mathrm{aux}}=
    \begin{pmatrix}
        {\mathbf{U}}^{\mathrm{aux}}_{\mathcal{L}_+\mathcal{L}_+} 
        &
        {\mathbf{U}}^{\mathrm{aux}}_{\mathcal{L}_+\mathcal{L}_-}
        &
        {\mathbf{U}}^{\mathrm{aux}}_{\mathcal{L}_+\mathcal{B}}\\
        {\mathbf{U}}^{\mathrm{aux}}_{\mathcal{L}_-\mathcal{L}_+} 
        &
        {\mathbf{U}}^{\mathrm{aux}}_{\mathcal{L}_-\mathcal{L}_-}
        &
        {\mathbf{U}}^{\mathrm{aux}}_{\mathcal{L}_-\mathcal{B}}\\
        {\mathbf{U}}^{\mathrm{aux}}_{\mathcal{B}\mathcal{L}_+} 
        &
        {\mathbf{U}}^{\mathrm{aux}}_{\mathcal{B}\mathcal{L}_-}
        &
        {\mathbf{U}}^{\mathrm{aux}}_{\mathcal{B}\mathcal{B}}
    \end{pmatrix}=
     \begin{pmatrix}
        0 
        &
        \mathbf{T}_{\mathcal{L}} \mathbf{U}_{\mathcal{L}\mathcal{L}}
        &
        \mathbf{T}_{\mathcal{L}} \mathbf{U}_{\mathcal{L}\mathcal{B}}\\
        \mathbf{T}_{\mathcal{L}}
        &
        0
        &
        0\\
        0 
        &
        \mathbf{U}_{\mathcal{B}\mathcal{L}}
        &
        \mathbf{U}_{\mathcal{B}\mathcal{B}}
        \end{pmatrix}
\end{equation}
where four blocks vanish due to the connectivity of the auxiliary 
graph, the other four blocks can been identified with corresponding blocks of the 
quantum map of the open graph and we introduced
\(\mathbf{T}_{\mathcal{L}}\equiv e^{ik_+ \mathbf{L}_{\mathcal{L}}}\), an \(N_{\mathcal{L}}\)-dimensional diagonal matrix that contains the  auxiliary lengths 
of the leads in the phase. Note, that \(\mathbf{T}_{\mathcal{L}} \to 0\)
as the auxiliary lengths are sent to infinity. 
Writing the identity \({\mathbf{U}}^{\mathrm{aux}}= 
\mathbf{R} - {\mathbf{U}}^{\mathrm{aux}}\mathbf{R} \)
in terms of its blocks one may express the blocks of \(\mathbf{R}\)
in the form
\begin{equation}
    \mathbf{R}=
    \begin{pmatrix}
{\mathbf{R}}_{\mathcal{L}_+\mathcal{L}_+} 
        &
{\mathbf{R}}_{\mathcal{L}_+\mathcal{L}_-}
        &
        {\mathbf{R}}_{\mathcal{L}_+\mathcal{B}}\\
        {\mathbf{R}}_{\mathcal{L}_-\mathcal{L}_+} 
        &
        {\mathbf{R}}_{\mathcal{L}_-\mathcal{L}_-}
        &
        {\mathbf{R}}_{\mathcal{L}_-\mathcal{B}}\\
        {\mathbf{R}}_{\mathcal{B}\mathcal{L}_+} 
        &
        {\mathbf{R}}_{\mathcal{B}\mathcal{L}_-}
        &
        {\mathbf{R}}_{\mathcal{B}\mathcal{B}}
    \end{pmatrix}=
     \begin{pmatrix}
     \mathbf{T}_{\mathcal{L}}\boldsymbol{\sigma}
        \frac{\mathbb{I}}{\mathbb{I}
        -\mathbf{T}_{\mathcal{L}}^2\boldsymbol{\sigma}}
        \mathbf{T}_{\mathcal{L}}
        &
        \mathbf{T}_{\mathcal{L}}\boldsymbol{\sigma}
        \frac{\mathbb{I}}{\mathbb{I}
        -\mathbf{T}_{\mathcal{L}}^2\boldsymbol{\sigma}}
        &
        \mathbf{T}_{\mathcal{L}}\frac{\mathbb{I}}{\mathbb{I}
        -\boldsymbol{\sigma}\mathbf{T}_{\mathcal{L}}^2}
        \boldsymbol{\rho}^{\mathrm{out}}
        \\
        \frac{\mathbb{I}}{\mathbb{I}
        -\mathbf{T}_{\mathcal{L}}^2\boldsymbol{\sigma}}\mathbf{T}_{\mathcal{L}} 
        &
        \frac{\mathbf{T}_{\mathcal{L}}^2\boldsymbol{\sigma}}{\mathbb{I}
        -\mathbf{T}_{\mathcal{L}}^2\boldsymbol{\sigma}}
        & 
        \mathbf{T}_{\mathcal{L}}^2\frac{\mathbb{I}}{\mathbb{I}
        -\boldsymbol{\sigma}\mathbf{T}_{\mathcal{L}}^2}
        \boldsymbol{\rho}^{\mathrm{out}}
        \\
        \boldsymbol{\rho}^\mathrm{in}
        \frac{\mathbb{I}}{\mathbb{I}
        -\mathbf{T}_{\mathcal{L}}^2\boldsymbol{\sigma}}\mathbf{T}_{\mathcal{L}} 
        &
        \boldsymbol{\rho}^\mathrm{in}
        \frac{\mathbb{I}}{\mathbb{I}
        -\mathbf{T}_{\mathcal{L}}^2\boldsymbol{\sigma}} 
        &
        \frac{\mathbf{U}_{\mathcal{BB}}}{\mathbb{I}- 
        \mathbf{U}_{\mathcal{BB}}}
    +\boldsymbol{\rho}^{\mathrm{in}}\mathbf{T}_{\mathcal{L}}^2\frac{\mathbb{I}}{\mathbb{I}
        -\boldsymbol{\sigma}\mathbf{T}_{\mathcal{L}}^2}
        \boldsymbol{\rho}^{\mathrm{out}}
        \end{pmatrix}
\end{equation}
where \(\boldsymbol{\sigma}\equiv
\mathbf{U}_{\mathcal{LL}} +\mathbf{U}_{\mathcal{LB}}
\frac{\mathbb{I}}{\mathbb{I} - \mathbf{U}_{\mathcal{BB}}}
\mathbf{U}_{\mathcal{BL}} \) is the scattering matrix of the
open graph, \(\boldsymbol{\rho}^{\mathrm{in}}=  
\frac{\mathbb{I}}{\mathbb{I}- \mathbf{U_{BB}}} \mathbf{U}_{\mathcal{BL}}\)  
and
\(\boldsymbol{\rho}^{\mathrm{out}}=  \mathbf{U}_{\mathcal{LB}}
\frac{\mathbb{I}}{\mathbb{I}- \mathbf{U_{BB}}} \).\\
To proceed one chooses
two points \(\mathbf{x}=(x_e,e)\) and \(\mathbf{x}'=(x_{e'},e')\) on 
the auxiliary graph \( \mathcal{G}^{\mathrm{aux}}\) and expresses
the Green's function 
\eqref{compact_Greens_function} of \( \mathcal{G}^{\mathrm{aux}}\)
in terms of appropriate matrix elements of \(\mathbf{R}\) and then
performs the limit \(\mathbf{T}_{\mathcal{L}}\to 0\).
Let us do this explicitly for \(e, e' \in \mathcal{L}\) and write 
\eqref{compact_Greens_function} for this case in the form
\begin{align}
    \label{aux_Greens_function}
    2 k_+ i \ G^{\mathrm{aux}}(\mathbf{x},\mathbf{x'}, E_+)= &
    \delta_{e,e'}\, e^{i k_+ |x_{e} -x'_{e'}|}+
     e^{i k_+(x_{e}-x'_{e'})} 
     \left[\mathbf{T}_{\mathcal{L}}^{-1}\mathbf{R}_{\mathcal{L}_+\mathcal{L}_+} \mathbf{T}_{\mathcal{L}}
     \right]_{ee'}\nonumber \\
     &+
     e^{-i k_+(x_{e}-x'_{e'})} 
     \left[\mathbf{R}_{\mathcal{L}_-\mathcal{L}_-} 
     \right]_{ee'}
     +
     e^{i k_+(x_{e}+x'_{e'})} 
     \left[\mathbf{T}_{\mathcal{L}}^{-1}\mathbf{R}_{\mathcal{L}_+\mathcal{L}_-} 
     \right]_{ee'}\nonumber\\
     &+
     e^{i k_+(x_{e}+x'_{e'})} 
     \left[\mathbf{R}_{\mathcal{L}_-\mathcal{L}_+}
     \mathbf{T}_{\mathcal{L}} 
     \right]_{ee'}    \nonumber \\
     = &
    \delta_{e,e'}\, e^{i k_+ |x_{e} -x'_{e'}|}+
     e^{i k_+(x_{e}-x'_{e'})} 
     \left[
        \frac{\boldsymbol{\sigma} \mathbf{T}_{\mathcal{L}}^2}{\mathbb{I}
        -\boldsymbol{\sigma}\mathbf{T}_{\mathcal{L}}^2}
     \right]_{ee'}\nonumber \\
     &+
     e^{-i k_+(x_{e}-x'_{e'})} 
     \left[\frac{\mathbf{T}_{\mathcal{L}}^2\boldsymbol{\sigma}}{\mathbb{I}
        -\mathbf{T}_{\mathcal{L}}^2\boldsymbol{\sigma}}
     \right]_{ee'}
     +
     e^{i k_+(x_{e}+x'_{e'})} 
     \left[\boldsymbol{\sigma}
        \frac{\mathbb{I}}{\mathbb{I}
        -\mathbf{T}_{\mathcal{L}}^2\boldsymbol{\sigma}}
     \right]_{ee'}\nonumber\\
     &+
     e^{i k_+(x_{e}+x'_{e'})} 
     \left[
     \frac{\mathbb{I}}{\mathbb{I}
        -\mathbf{T}_{\mathcal{L}}^2\boldsymbol{\sigma}}\mathbf{T}_{\mathcal{L}}^2  
     \right]_{ee'}   
\end{align}
where we may now send the edge lengths of the leads to infinity
\(\mathbf{T}_{\mathcal{L}}\to 0\). This results in 
\begin{equation}
    2 k_+ i \ G(\mathbf{x},\mathbf{x'}, E_+)=
     \delta_{e,e'}\, e^{i k_+ |x_{e} -x'_{e'}|}+
     e^{i k_+(x_{e}+x'_{e'})} 
     \boldsymbol{\sigma}_{ee'}
\end{equation}
which is equivalent to the given expression for the open 
Green's function \eqref{open_Greens_function} if both points are on the leads.
The other cases can be derived in the same way. 
This calculation is equivalent to formally expanding the Green's function of the 
auxiliary graph as a sum over trajectories. Sending the lengths of the leads to 
infinity is equivalent to only summing over trajectories that never travel through
any lead from one end to the other - summing just these trajectories then
gives back \eqref{open_Greens_function}.

\section{Regularity of the scattering matrix $\boldsymbol{\sigma}$ at a bound 
state in the continuum} \label{AppB}
Following on from the discussion in Sec.~\ref{regularisation}, we 
show here that the singularity of the scattering matrix 
\(\boldsymbol{\sigma}(k)\) and the coupling matrix 
\(\boldsymbol{\rho}(k)\), Eqs. \eqref{scattering_matrix}
and\eqref{scattering_state}, in the presence of a perfect scar (described by 
the eigenvector \(\mathbf{b}_0\)) can be lifted and that the solution is regular 
across a whole \(k\) interval containing \(k_0\).

\subsection{Closed expressions for \(\mathbf{P}\boldsymbol{\rho}(k)\)}

First, we decompose the internal graph amplitudes of a scattering solution 
(\ref{rho_definition}), that is, \(\mathbf{a}
(k)^{\mathrm{in}}_{\mathcal{B}}=\boldsymbol{\rho}
(k)\mathbf{a}_{\mathcal{L}}^{\mathrm{in}}\), into components parallel and 
orthogonal to \(\mathbf{b}_{0}\), 
\begin{equation}
    \mathbf{P}\, \mathbf{a}(k)^{\mathrm{in}}_{\mathcal{B}} +
    \mathbf{Q} \, \mathbf{a}(k)^{\mathrm{in}}_{\mathcal{B}}= (\mathbf{P} 
    \boldsymbol{\rho}(k)+ 
    \mathbf{Q}\boldsymbol{\rho}(k))\, \mathbf{a}_{\mathcal{L}}^{\mathrm{in}},
\end{equation}
where the projection operator and its orthogonal component are defined in 
(\ref{projectionP}) and (\ref{projectionQ}). Starting from Eq.\ (\ref{internal}), 
we write 
\begin{eqnarray*}
 \mathbf{P} \left(\mathbb{I}-\mathbf{U}(k)_{\mathcal{BB}}\right)
 (\mathbf{P}+\mathbf{Q})\,\mathbf{a}^{\mathrm{in}}_{\mathcal{B}} &=& 
   \mathbf{P} \mathbf{U}(k)_{\mathcal{BL}}\, 
   \mathbf{a}^{\mathrm{in}}_{\mathcal{L}},\\
   \mathbf{Q} \left(\mathbb{I}-\mathbf{U}(k)_{\mathcal{BB}}\right)
   (\mathbf{P}+\mathbf{Q})\,\mathbf{a}^{\mathrm{in}}_{\mathcal{B}} &=& 
   \mathbf{Q} \mathbf{U}(k)_{\mathcal{BL}}\,\mathbf{a}^{\mathrm{in}}_{\mathcal{L}},
\end{eqnarray*}
which yields
\begin{subequations}
\begin{eqnarray} \label{rhoP0}
 \left(
   {\mathbf{b}_{\mathcal{B}}^\mathrm{in}}^{\dagger}
    \left(
    \mathbb{I}-\mathbf{U}(k)_{\mathcal{BB}}
    \right)
    \mathbf{b}_{\mathcal{B}}^\mathrm{in}\right)
    \cdot \mathbf{P} \, \mathbf{a}^{\mathrm{in}}_{\mathcal{B}} - \mathbf{P} 
    \mathbf{U}(k)_{\mathcal{BB}} \mathbf{Q}\, 
    \mathbf{a}^{\mathrm{in}}_{\mathcal{B}} &=& 
   \mathbf{P} \mathbf{U}(k)_{\mathcal{BL}}\, 
   \mathbf{a}^{\mathrm{in}}_{\mathcal{L}},\\ \label{rhoQ0}
   - \mathbf{Q} \mathbf{U}(k)_{\mathcal{BB}} \mathbf{P} 
   \,\mathbf{a}^{\mathrm{in}}_{\mathcal{B}} + \mathbf{Y}_Q(k) 
   \mathbf{Q}\,\mathbf{a}^{\mathrm{in}}_{\mathcal{B}} &=& 
   \mathbf{Q} \mathbf{U}(k)_{\mathcal{BL}}\, 
   \mathbf{a}^{\mathrm{in}}_{\mathcal{L}},
\end{eqnarray}
\end{subequations}
where \(\mathbf{Y}_Q(k)\) 
has been defined in \eqref{YQ} 
We have defined \(\mathbf{Y}_Q(k)^{-1}\) 
in \eqref{YQ-inverse-definition}
as the inverse on the reduced space spanned by \(\mathbf{Q}\).
Note that these definitions are here extended to wave numbers close to
\(k_0\) while \(\mathbf{P}\) and \(\mathbf{Q}\) do not depend on \(k\).
We used the general relation \(\mathbf{P}\mathbf{A} 
\mathbf{P}= \left({\mathbf{b}_{\mathcal{B}}^\mathrm{in}}^{\dagger}
\mathbf{A}\mathbf{b}_{\mathcal{B}}^\mathrm{in}\right)
\cdot \mathbf{P}\) for a square 
matrix \(\mathbf{A}\). 
After rearranging (\ref{rhoQ0}) by multiplying with 
\(\mathbf{Y}_Q(k)^{-1}\) and replacing
\(\mathbf{a}(k)^{\mathrm{in}}_{\mathcal{B}}\) by 
\(\boldsymbol{\rho}(k)\,\mathbf{a}_{\mathcal{L}}^{\mathrm{in}}\), we obtain
\begin{equation}
   \mathbf{Q} \boldsymbol{\rho}(k) =
    \mathbf{Y}_Q(k)^{-1} \mathbf{U}(k)_{\mathcal{BB}}
    \mathbf{P} \boldsymbol{\rho}(k) +
    \mathbf{Y}_Q(k)^{-1} \mathbf{U}(k)_{\mathcal{BL}}\,.
    \label{Qrho}
\end{equation}
Given that \(
{\mathbf{b}_{\mathcal{B}}^\mathrm{in}}^{\dagger}
\left(\mathbb{I}-\mathbf{U}(k)_{\mathcal{BB}}\right)
\mathbf{b}_{\mathcal{B}}^\mathrm{in}\) in 
(\ref{rhoP0}) is a scalar and after replacing \(\mathbf{Q}\, 
\mathbf{a}^{\mathrm{in}}_{\mathcal{B}} \) by 
\(\mathbf{Q}\boldsymbol{\rho}(k)\,  
\mathbf{a}^{\mathrm{in}}_{\mathcal{L}}\) using (\ref{Qrho}), one 
obtains after some 
further manipulations
\begin{equation}
    \mathbf{P}\boldsymbol{\rho}(k)=
    \mathbf{P} \frac{\mathbb{I}+ \mathbf{U}(k)_{\mathcal{BB}} 
    \mathbf{Y}_Q(k)^{-1}  }{
    \mathbf{b}^{\mathrm{in}\, \dagger}_{\mathcal{B}}\left[
    \mathbb{I}- \mathbf{U}(k)_{\mathcal{BB}} -
    \mathbf{U}(k)_{\mathcal{BB}}
    \mathbf{Y}_Q(k)^{-1}\mathbf{U}(k)_{\mathcal{BB}}
    \right] 
    \mathbf{b}^{\mathrm{in}}_{\mathcal{B}}}
    \mathbf{U}(k)_{\mathcal{BL}}.
    \label{Prho}
\end{equation}
In order to analyse the scattering solutions in the vicinity of 
the bound state, we consider
wave numbers \(k\) close to \(k_0\) in the limit
\(\delta k \equiv k-k_0 \to 0\) in 
the matrices \(\boldsymbol{\sigma}(k)\) 
and \(\boldsymbol{\rho}(k)\).
By construction we have 
\(\mathbf{Y}_Q(k) \mathbf{b}_{\mathcal{B}}^\mathrm{in} =
0 \) 
and \( \mathbf{Y}_Q(k)^{-1}\) has been defined 
on the subspace spanned by the projector \(\mathbf{Q}\) in 
order to remove the 
pole at \(k_0\).
For wave numbers \(k\) sufficiently close to \(k_0\) 
this definition remains well defined 
due to the (assumed) non-degeneracy of \(\mathbf{U}(k)\) as
the matrix is then free of poles.

\subsection{Expansion of \(\mathbf{P}
\boldsymbol{\rho}(k)\) around \(k=k_0\)}\label{sec: Approaching 
scared states}

We will show in the following that, as \(k \to k_0\) in 
\eqref{Prho},
the denominator 
\( \mathbf{b}^{\mathrm{in}\, \dagger}_{\mathcal{B}}\left[
\mathbb{I}- \mathbf{U}(k)_{\mathcal{BB}}
-\mathbf{U}(k)_{\mathcal{BB}}
\mathbf{Y}_Q(k)^{-1}\mathbf{U}(k)_{\mathcal{BB}}
\right] 
\mathbf{b}^{\mathrm{in}}_{\mathcal{B}}\) vanishes but so 
does the numerator. We will show this for vertex 
scattering matrices of the 
form (\ref{general_vertex_scattering}) by performing a 
Taylor expansion of both expressions around \(k=k_0\). For 
this, we need to find explicit expressions for the 
derivative of the blocks of the quantum map 
\(\mathbf{U}(k)\). 
The calculation of these is similar to the one performed in 
\ref{normalization-derivation} using Eq.~\eqref{scattering derivative}.
When this equation is applied here to the full 
quantum map $\mathbf U$, one obtains
\begin{equation}
    \frac{d}{dk} \mathbf{U}(k)=
    \begin{pmatrix}
    0 & 0\\
    0 & i \boldsymbol{\mathbf{L}}
    \end{pmatrix} \mathbf{U}(k)
    +
    \frac{1}{2k}
    \left[
    \begin{pmatrix}
    \mathbb{I} & 0\\
    0 & e^{ik \mathbf{L}} \boldsymbol{\Pi}
    \end{pmatrix}
     - \mathbf{U}(k)  
     \begin{pmatrix}
    \mathbb{I} & 0\\
    0 & e^{-ik\mathbf{L}} \boldsymbol{\Pi}
    \end{pmatrix}
     \mathbf{U}(k)
 \right]\ ,
\end{equation}
where $\mathbf{L}$ and $\exp(-ik\mathbf{L})$ are $2 {N}_B$-
dimensional diagonal 
matrices with diagonal entries $\ell_e$ and $\exp(-ik\ell_e)$, 
respectively. 
Setting \(k= k_0 +\delta k\), we find the expansions
\begin{subequations}
\begin{align}
    \mathbf{U}(k_0+\delta k)_{\mathcal{BB}}=
    &
    \mathbf{U}(k_0)_{\mathcal{BB}}+ i \delta k 
    \mathbf{L} \mathbf{U}(k_0)_{\mathcal{BB}} \nonumber\\
    &+\frac{\delta k}{2k_0}
    \left(e^{ik_0 \mathbf{L}}\boldsymbol{\Pi} - 
    \mathbf{U}(k_0)_{\mathcal{BB}}e^{-ik_0 
    \mathbf{L}}\boldsymbol{\Pi} 
    \mathbf{U}(k_0)_{\mathcal{BB}}\right) \nonumber\\
    &
    -\frac{\delta k}{2k_0}  \mathbf{U}(k_0)_{\mathcal{BL}}
    \mathbf{U}(k_0)_{LB}
    + O((\delta k)^2)\\
    \mathbf{U}(k_0+\delta k)_{\mathcal{BL}}=&
    \mathbf{U}(k_0)_{\mathcal{BL}}+  i \delta k 
    \mathbf{L} \mathbf{U}(k_0)_{\mathcal{BL}} \nonumber\\
    &-\frac{\delta k}{2 k_0}\mathbf{U}(k_0)_{\mathcal{BB}}
    e^{-ik_0\mathbf{L}}\boldsymbol{\Pi} 
    \mathbf{U}(k_0)_{\mathcal{BL}}
    \nonumber\\ 
    &- \frac{\delta k}{2 k_0} \mathbf{U}(k_0)_{\mathcal{BL}}
    \mathbf{U}(k_0)_{\mathcal{LL}} + O((\delta k)^2)\ .
\end{align}    
\end{subequations}
As \(\mathbf{b}^{\mathrm{in}}_{\mathcal{B}}\) is a normalized eigenvector
of \(\mathbf{U}(k_0)_{\mathcal{BB}}\) with eigenvalue one and as 
\( \mathbf{U}(k_0)_{\mathcal{LB}} \mathbf{b}^{\mathrm{in}}_{\mathcal{B}}=0\), 
\({\mathbf{b}^{\mathrm{in}}_{\mathcal{B}}}^{\dagger} 
\mathbf{U}(k_0)_{\mathcal{BL}}=0\) due to the unitarity of
\(\mathbf{U}(k_0)\), one gets
\begin{equation}
  {\mathbf{b}^{\mathrm{in}}_{\mathcal{B}}}^{\dagger}  
  \mathbf{U}(k_0+\delta k)_{\mathcal{BB}}
  \mathbf{b}^{\mathrm{in}}_{\mathcal{B}}
  = 1 +i \delta k 
  {\mathbf{b}^{\mathrm{in}}_{\mathcal{B}}}^{\dagger}  
  \left(\mathbf{L} + \frac{\sin(k_0\mathbf{L})}{k_0} \boldsymbol{\Pi}\right)
  \mathbf{b}^{\mathrm{in}}_{\mathcal{B}} + O((\delta k)^2)
\end{equation}
and
\begin{equation}
  {\mathbf{b}^{\mathrm{in}}_{\mathcal{B}}}^{\dagger}  
  \mathbf{U}(k_0+\delta k)_{\mathcal{BB}}\mathbf{Y}_Q(k_0+\delta k)^{-1} 
  \mathbf{U}(k_0+\delta k)_{BB}
  \mathbf{b}^{\mathrm{in}}_{\mathcal{B}}
  = O((\delta k)^2).
\end{equation}
The last two equations together give
\begin{multline}
    {\mathbf{b}^{\mathrm{in}}_{\mathcal{B}}}^{\dagger}\left[
\mathbb{I}- \mathbf{U}(k)_{\mathcal{BB}}
-\mathbf{U}(k)_{\mathcal{BB}}\mathbf{Y}_Q(k)^{-1}\mathbf{U}(k)_{\mathcal{BB}}
\right] 
\mathbf{b}^{\mathrm{in}}_{\mathcal{B}}\\ 
 =\ -i \delta k\ {\mathbf{b}^{\mathrm{in}}_{\mathcal{B}}}^{\dagger}
 \left[ \mathbf{L}
    +  \frac{\sin(\mathbf{L} k_0 )}{k_0} \boldsymbol{\Pi} \right] 
    \mathbf{b}^{\mathrm{in}}_{\mathcal{B}}
    \, +\, O((\delta k)^2)\, .
    \label{Prho_denominator}
\end{multline}

Analogously one finds
\begin{equation}
    \mathbf{P} \mathbf{U}(k_0+\delta k)_{\mathcal{BL}}=
    i \mathbf{P} \mathbf{L}  \mathbf{U}(k_0)_{\mathcal{BL}} \delta k
    -\mathbf{P} \frac{\delta k}{2 k_0}e^{-ik_0\mathbf{L}}\boldsymbol{\Pi} 
    \mathbf{U}(k_0)_{\mathcal{BL}}
    +O((\delta k)^2)
\end{equation}
and
\begin{multline}
  \mathbf{P} \mathbf{U}(k_0+\delta k)_{\mathcal{BB}}\mathbf{Q}=\\
  \delta k\  \mathbf{P}
  \left[
  i  
  \mathbf{L} \mathbf{U}(k_0)_{\mathcal{BB}}
  +\frac{1}{2k_0}\boldsymbol{\Pi} 
  \left(e^{ik_0\mathbf{L}} - e^{-ik_0\mathbf{L}} 
  \mathbf{U}(k_0)_{\mathcal{BB}}\right)
  \right] \mathbf{Q}
  + O((\delta k)^2)
\end{multline}
which together yield 
\begin{multline}
    \mathbf{P}(\mathbb{I} + 
    \mathbf{U}(k)_{\mathcal{BB}}\mathbf{Y}_Q(k)^{-1})
    \mathbf{U}(k)_{\mathcal{BL}} =\\
     i \delta k\ \mathbf{P} \left[
    \mathbf{L} - \frac{1}{2 k_0 i}\boldsymbol{\Pi} e^{-i k_0 \mathbf{L} }  
    \right] \mathbf{U}(k_0)_{\mathcal{BL}}\\
    +
     i \delta k\ \mathbf{P} 
    \left[
    \left( \mathbf{L} \mathbf{U}(k_0)_{\mathcal{BB}} + \boldsymbol{\Pi} 
    \frac{e^{i k_0 \mathbf{L}}- e^{-ik_0\mathbf{L}}
    \mathbf{U}(k_0)_{\mathcal{BB}}}{2 k_0 i}
    \right) \mathbf{Y}_Q(k_0)^{-1}
    \right]\mathbf{U}(k_0)_{\mathcal{BL}} + O((\delta k)^2)\ .
    \label{Prho_enumerator}
\end{multline}

Finally, we show that the term 
\({\mathbf{b}^{\mathrm{in}}_{\mathcal{B}}}^{\dagger}  
\left(\mathbf{L} +\frac{1}{k_0}\sin(k_0\mathbf{L})\boldsymbol{\Pi}  \right)
\mathbf{b}^{\mathrm{in}}_{\mathcal{B}}\) in (\ref{Prho_denominator}) does not 
vanish. This is essential 
for the limit  \( \lim_{\delta k\to 0 } \mathbf{P} \boldsymbol{\rho}(k+\delta k) 
\) to be well defined (and finite). Indeed one has
\begin{equation}
{\mathbf{b}^{\mathrm{in}}_{\mathcal{B}}}^{\dagger}  
  \left(\mathbf{L} + \frac{\sin(k_0\mathbf{L})}{k_0}\boldsymbol{\Pi} \right)
  \mathbf{b}^{\mathrm{in}}_{\mathcal{B}}=
  \sum_{e \in \mathcal{B}}
  \ell_e(|b_{e_+}|^2
  +|b_{e_-}|^2)
  + \frac{\sin(k_0 \ell_e)}{k_0} \left(
  b_{e_+}^{ *}b_{e_-}+
  b_{e_-}^{*}b_{e_+}
  \right)
\end{equation}
which is a sum over positive terms as (for \(k_0>0\))
\begin{multline*}
\left|
\frac{\sin(k_0 \ell_e)}{k_0 \ell_e} 
\left(
b_{e_+}^{ *}b_{e_-}+
b_{e_-}^{ *}b_{e_+}
\right)
\right|< 
\left| 
\left(
b_{e_+}^{*}b_{e_-}+
b_{e_-}^{ *}b_{e_+}
\right)
\right|\le |b_{e_+}|^2+|b_{e_-}|^2
\end{multline*}
using the Cauchy-Schwartz inequality. \\

This means that the limit \(\mathbf{P}\boldsymbol{\rho}(k_0)\equiv 
\lim_{\delta k \to 0} \mathbf{P}\boldsymbol{\rho}(k_0 + \delta k)\) is well defined and we obtain to 
leading order 
\begin{equation}
    \mathbf{P} 
    \boldsymbol{\rho}(k_0) =
    \frac{\mathbf{P}
     \left[
     \frac{1}{2i}\boldsymbol{\Pi} e^{-i k_0 \mathbf{L} }  
     -k_0 \mathbf{L}
    -\left(
    k_0 \mathbf{L} \mathbf{U}_{\mathcal{BB}} 
    + \boldsymbol{\Pi} 
    \frac{e^{i k_0 \mathbf{L}} -
    e^{-ik_0 \mathbf{L}}\mathbf{U}_{\mathcal{BB}}}{2i}
    \right) \mathbf{Y}_Q^{-1}
    \right]
    }{ 
    {\mathbf{b}^{\mathrm{in}}_{\mathcal{B}}}^{\dagger}\left[k_0 \mathbf{L}
    +  \sin(k_0 \mathbf{L})\boldsymbol{\Pi} 
    \right] 
    \mathbf{b}^{\mathrm{in}}_{\mathcal{B}}
    }
    \mathbf{U}(k_0)_{\mathcal{BL}}.
\end{equation}
For quantum graphs with vertex matching conditions leading to
vertex scattering matrices not depending on the wave number, (such as Neumann-
Kirchhoff boundary conditions),
this simplifies further to
\begin{equation}
    \mathbf{P} \boldsymbol{\rho}(k_0)  =-\mathbf{P} \mathbf{L}\
    \frac{
      \mathbb{I}   
    + \mathbf{U}(k_0)_{\mathcal{BB}} \mathbf{Y}_Q(k_0)^{-1}
    }{ 
    {\mathbf{b}^{\mathrm{in}}_{\mathcal{B}}}^{\dagger} \mathbf{L}\
     \mathbf{b}^{\mathrm{in}}_{\mathcal{B}}}
    \mathbf{U}(k_0)_{\mathcal{BL}}
    \ .
\end{equation}
Likewise, it can be shown that \(\mathbf{Q}\boldsymbol{\rho}\) 
in \eqref{Qrho} 
and the scattering matrix in \eqref{sigma_in_terms_of_rho} 
are also well defined in an interval containing \(k_0\). In the limit $k\to k_0$, 
we obtain for the latter the result (\ref{s-matrix-reg}) as expected.

In this regularization, we have explicitly used Eq.\ \eqref{scattering derivative}
which is valid precisely for scattering matrices that come from a self-adjoint 
matching condition. So one may wonder whether it is valid for the large amount of 
physical quantum graph models that define the quantum graph in terms of arbitrary 
prescribed scattering matrices (as for instance in \cite{BG01}). In most of these 
physical cases, the scattering matrices are assumed to be constant with respect to
\(k\) which implies that the right-hand side of Eq.\ \eqref{scattering derivative}
vanishes. It is easy to see that this leads to some 
simplifications in the following formulas and leads to a 
well-defined regularized scattering matrix. If one prescribes scattering matrices 
with some dependency on the wave number then the regularity of the scattering
matrices in the presence of bound states cannot be guaranteed in general. However
if the scattering matrix is an effective description derived from a more detailed
self-adjoint system (whether that is a graph or a different type of model), 
then there exists a well-defined scattering matrix 
both physically and mathematically basically because the spectral 
decomposition of self-adjoint 
operators is always based on orthogonal projections, such that scattering 
states are 
always orthogonal to bound states. Showing the regularity in this case will 
require an analogous projection method but will generally require its own 
analysis. Vice versa a non-regular scattering matrix may be an indicator that
a model is not physical in all respects (which does not necessarily mean that
the model is bad as long as its limitations are known).

Our assumption that the perfect scar is non-degenerate may also be lifted but 
leads to more cumbersome calculations -- if the perfect scars 
do not overlap, one may regularise by
first regularizing the scattering matrices of the corresponding non-overlapping
subgraphs and then build up the full scattering matrix from there. Otherwise the 
rank one projector \(\mathbf{P}\) needs to be replaced by higher rank projectors.

\bibliographystyle{iopart-num}
\bibliography{bibliography}
\end{document}